\documentclass[aps,prb,amsmath,amssymb,citeautoscript,reprint]{revtex4-2}
\usepackage{graphicx, float, xcolor, pgf}
\usepackage{physics, siunitx}
\usepackage[caption=false, justification=centerlast]{subfig}

\makeatletter
\let\oldtheequation\theequation
\renewcommand\tagform@[1]{\maketag@@@{\ignorespaces#1\unskip\@@italiccorr}}
\renewcommand\theequation{(\oldtheequation)}
\makeatother

\usepackage{hyperref}
\hypersetup{
  pdfauthor={Amartya Bose},
  pdftitle={Tensor Network Path Integral},
  colorlinks=true,
  linkcolor=black,
  citecolor=black,
  urlcolor=black
}

\begin{document}
\title{Tensor Network Path Integrals: Implementation and Analysis}
\author{Amartya Bose}
\thanks{\label{ft:equalcontri}Both authors contributed equally to this work.}
\affiliation{Department of Chemistry, Princeton University, Princeton, New Jersey 08544}
\author{Peter L. Walters}
\thanks{\label{ft:equalcontri}Both authors contributed equally to this work.}
\affiliation{Department of Chemistry, University of California Berkeley, Berkeley, California 94720}
\affiliation{Miller Institute for Basic Research in Science, University of California Berkeley, Berkeley, California 94720}
%\affiliation{Department of Chemistry, University of California, Berkeley, California 94720}
\allowdisplaybreaks

\begin{abstract}
  Tensors with finite correlation afford very compact tensor network
  representations. A novel tensor network-based decomposition of real-time path
  integral simulations involving Feynman-Vernon influence functional is
  introduced. In this tensor network path integral (TNPI) technique, the finite
  temporarily non-local interactions introduced by the influence functional can
  be captured very efficiently using matrix product state representation for the
  path amplitude (PA) tensor. We illustrate this particular TNPI method through
  various realistic examples, including a charge transfer reaction and an
  exciton transfer in a dimer. We also show how it is readily applied to systems
  with greater than two states by simulating a 7-site model of FMO and a
  molecular wire model. The augmented propagator (AP) TNPI utilizes the
  symmetries of the problem, leading to accelerated convergence and dramatic
  reductions of computational effort. We also introduce an approximate method
  that speeds up propagation beyond the non-local memory length. Furthermore,
  the structure imposed by the tensor network representation of the PA tensor
  naturally suggests other factorizations that make simulations for extended
  systems more efficient. These factorizations would be the subject of future
  explorations. The flexibility of the AP-TNPI framework makes it a promising
  new addition to the family of path integral methods for non-equilibrium
  quantum dynamics.
\end{abstract}
\maketitle

\section{Introduction}
Quantum mechanics plays an important role in various condensed phase dynamical
processes. While there are certain processes, such as calculation of vibrational
spectra, that can be simulated using classical mechanics~\cite{Zhang2020},
dynamical problems that involve deep tunneling through potential barriers like
charge or excitonic transfer are essentially quantum mechanical in nature.
Simulation of these quantum processes in condensed phase environments is an
extremely challenging task owing to the exponential growth of computational
complexity with the degrees of freedom of the system being studied. Wave
function-based methods like multi-configuration time-dependent Hartree
(MCTDH)~\cite{Beck2000}, its multi-layer version (ML-MCTDH)~\cite{Wang2003}, and
time-dependent density matrix renormalization group
(tDMRG)~\cite{White2004,Schollwock2005,Ma2018}, for example, are extremely
powerful in efficiently simulating the dynamics of certain classes of large
quantum systems. Tremendous progress has been made in using these wave
function-based methods for simulating systems coupled to many degrees of
freedom~\cite{Maiolo2021,Picconi2019,Jiang2020,Ren2018}. However, if low
frequency ro-vibrational and translational degrees of freedom are involved in
the dynamics, the number of extra degrees of freedom that need to be considered
explicitly increases, increasing the computational requirements especially at
high temperatures. Typically, in the most general case, wave function-based
methods are also unable to rigorously describe the thermal equilibrium of
systems. These reasons can often render them ill-suited for studying the
dynamics of thermal condensed phase chemical systems when a solvent is
involved~\cite{Tanimura2020}.

Over the years, there has been an explosion of methods that attempt to solve the
problem of thermal quantum dynamics in a condensed phase. Most of these methods
can be categorized into two groups: approximate methods or quantum-classical
methods. Approximate methods tend to treat the entire system under study using
``augmented'' classical trajectories. The most prominent methods of this group
are the various approaches based on the semiclassical
approximation~\cite{Liu2006b,Liu2007,Gelabert2000,Gelabert2001,Guallar1999},
centroid molecular dynamics~\cite{Cao1994,Cao1994a} and ring polymer molecular
dynamics~\cite{Craig2004,Habershon2007,Rossi2014} for Born-Oppenheimer dynamics,
and the family of surface hopping
methods~\cite{Tully1971,Sholl1998,Tully2012,Miao2019} for non-adiabatic
processes. The quantum-classical methods, on the other hand, typically identify
a reduced-dimensional quantum ``system'' and relegate the rest of the degrees of
freedom to a classical treatment, thereby effectively restricting the
exponential growth of complexity of quantum dynamics to only the system degrees
of freedom. Various rigorous quantum-classical approaches have been developed
over the years such as quantum-classical Liouville dynamics~\cite{Kapral1999}
and quantum-classical path integral
(QCPI)~\cite{Lambert2012,Lambert2012a,Banerjee2013a,Walters2016}. Because the
system is now treated using exact quantum mechanics all effects stemming from
non-adiabaticity are rigorously accounted for.

Amongst the class of problems that can be solved using the system-solvent
decomposition, if the solvent can be modelled as a bath of harmonic oscillators,
the reduced density matrix of the system can be rigorously simulated without
classical trajectories using hierarchical equations of motion
(HEOM)~\cite{Tanimura1989} and the Feynman-Vernon influence functional
(IF)~\cite{Feynman1963}. Though general condensed phase solvents are anharmonic,
over the years, the harmonic bath mapping, obtained via linear response theory,
has repeatedly been shown to give impressively accurate results for a variety of
systems~\cite{Makri1999}. While HEOM is, in principle, a completely general
method for simulating these systems, in practice, significant challenges arise
while simulating the dynamics of a system coupled to a bath that is not
described by a Drude spectral density. Recent work has attempted at increasing
the applicability of HEOM to more general
cases~\cite{Tanimura1991,Popescu2016,Duan2017,Ikeda2020,Tanimura2020}. The
quasi-adiabatic propagator path integral (QuAPI)~\cite{Makri1995_1,Makri1995_2}
and methods based on QuAPI~\cite{Makri2014,Makri2014a,Makri2020,Makri2020a}
enable efficient simulations of the dynamics of systems coupled bi-linearly to
harmonic baths. Methods related to QuAPI and HEOM simulate the reduced density
matrix. So, the cost does not scale with the number of bath degrees of freedom,
and they are able to describe thermal equilibrium accurately. These methods
typically model the dynamics of open systems in cases where the system and the
solvent have different natures, e.g. a spin system coupled to a reservoir of
harmonic bath, and can easily be distinguished. Recently,~\citet{Lerose2021a}
have developed an influence functional approach for simulating cases where the
system and the bath do not have different natures, like identifying one
particular spin in a chain of spins as the system.

The biggest challenge in evaluating the real-time path integral is the non-local
nature of the IF. This non-Markovian memory leads to an exponential growth of
the number of paths that need to be tracked. However, it has a finite length,
which means that the entanglement between time points along a path decays after
a certain temporal distance. Tensors with finite entanglement can, on many
occasions, be very compactly represented using tensor networks. For instance,
with extended quantum systems, the dimensionality of the Hilbert space grows
exponentially with the number of particles. However, matrix-product states
(MPSs) and other tensor network decompositions allow efficient representations
of many such systems. This is why TN structures are used to efficiently
represent the wave function in
DMRG~\cite{White1992,Schollwock2005,Schollwock2011,Schollwock2011b,Stoudenmire2012,Stoudenmire2012a}
and other methods optimized for simulating multiple dimensions and critical
phenomena like the projected entangled-pair state (PEPS)~\cite{Orus2014} and the
multiscale entanglement renormalization ansatz
(MERA)~\cite{Vidal2007b,Vidal2008}. Of late, using tensor network algorithms for
simulating open quantum systems has also been gaining
popularity~\cite{Strathearn2018,Jorgensen2019,Lerose2021a,Ye2021,Bose2021PCTNPI}.
In this paper, we introduce and critically evaluate a similar tensor-network
method for performing path integral simulations for systems coupled with a
dissipative medium using the QuAPI splitting~\cite{Makri1992,Makri1993a}. This
method, the augmented propagator-based tensor network path integral (AP-TNPI),
uses MPSs to compress the representation of the path amplitude (PA) tensor
involved in path integral simulations. Consequently, the influence functional
becomes a matrix-product operator (MPO) that acts on the PA MPS. As opposed to
the other numerical approaches for defining these matrix product
structures~\cite{Strathearn2018,Ye2021,Cygorek2021}, by a clean separation of
the path amplitude tensor from the influence functional MPO and through a
careful analytical construction of the latter, we exploit inherent symmetries
present in the problem giving closed form expressions to it. In this work, we
utilize the open source ITensor library~\cite{ITensor} as the basis of our
AP-TNPI implementation, as it provides for an efficient means of evaluating the
relevant tensor operations.

In Sec.~\ref{sec:method}, we discuss the AP-TNPI method, detailing a
construction that results in a simple, elegant and efficient structure. We also
demonstrate an approximation that emerges naturally from the resulting
structure. This ``Markovian iteration scheme'' approximation is able to speed up
simulations involving processes with very long time-scales and comparatively
short memory lengths. Thereafter, in Sec.~\ref{sec:result}, we illustrate the
use of the method through numerous examples. Apart from a direct comparison with
QuAPI and related methods, we apply AP-TNPI to study charge and exciton transfer
dynamics. We demonstrate the ability of AP-TNPI to develop an optimized
representation enabling us to access long memory times even with very modest
computational resources. Finally, we end the paper in Sec.~\ref{sec:conclusion}
with some concluding remarks, and outlook for further developments based on, and
facilitated by AP-TNPI.

\section{Methodology}\label{sec:method}
Consider a potentially time-dependent quantum system interacting bi-linearly with
a harmonic bath. The Hamiltonian is given as:
\begin{align}
  \hat{H}                                                                    & = \hat{H}_{0}(t) + \hat{H}_{\text{b}}\left(\mathbf{\hat{x}}, \mathbf{\hat{p}}, \hat{s}\right)                                                                     \\
  \hat{H}_{\text{b}}\left(\mathbf{\hat{x}}, \mathbf{\hat{p}}, \hat{s}\right) & = \sum_{j} \frac{\hat{p}_{j}^{2}}{2m_{j}} + \frac{1}{2}m_{j}\omega_{j}^{2}\left(\hat{x}_{j} - \frac{c_{j}\hat{s}}{m_{j}\omega_{j}^{2}}\right)^{2}\label{eq:Hamil}
\end{align}
where $\hat{H}_{0}(t)$ is the Hamiltonian of the isolated $D$-state quantum
system, and $\hat{H}_{\text{b}}$ is the Hamiltonian for the bath and the
system-bath interactions. These interactions are defined in terms of $\omega_{j}$ and
$c_{j}$, which are the frequency and the coupling strength of the
$j$\textsuperscript{th} bath mode, respectively. The system position operator is
$\hat{s}$. The system-bath interaction is characterized by the spectral
density~\cite{Caldeira1983}:
\begin{align}
  J(\omega) & = \frac{\pi}{2}\sum_{j}\frac{c_{j}^{2}}{m_{j}\omega_{j}}\delta\left(\omega - \omega_{j}\right).
\end{align}
The spectral density is related to the energy gap autocorrelation function via a
Fourier transform~\cite{Makri1999}. For most classical solvents, this
autocorrelation function is computing using classical trajectories on an
appropriately parameterized force field.

The dynamics of the isolated $D$-state system can be obtained by directly
solving the time-dependent Schr\"{o}dinger equation for the propagator:
\begin{align}
  i\hbar \frac{\partial\hat{U}_{0}(t_{0},t)}{\partial t} & = \hat{H}_{0}(t) \hat{U}_{0}(t_{0},t),
\end{align}
where $\hat{U}_{0}(t_{0},t)$ is the propagator evolving the system from an
initial state at time $t_{0}$ to a final state at time $t$. The density matrix
is propagated using a combination of forward and backward propagators; this
combined forward-backward propagator, $K_{j}$, connecting states at time point
$t_{0}=(j-1)\Delta t$ to $t=j\Delta t$ is given by:
\begin{align}
  K_{j}\left(s_{j-1}^{\pm},s_{j}^{\pm},\Delta t\right) & =\mel{s_{j}^{+}}{\hat{U}_{0}((j-1)\Delta t,j\Delta t)}{s_{j-1}^{+}}\nonumber                       \\
                                                       & \times \mel{s_{j-1}^{-}}{\hat{U}^{\dag}_{0}((j-1)\Delta t,j\Delta t)}{s_{j}^{-}}\label{eq:FB_Prop}
\end{align}
where $s_{j}^{\pm}$ is the forward-backward state of the system
at the $j$\textsuperscript{th} time point discretized according to QuAPI
splitting~\cite{Makri1992,Makri1993a}.

Suppose that the system and bath are initially uncoupled and that the bath is in
a thermal equilibrium $\rho_{\text{b}}(\beta)$ at an inverse temperature of $\beta$. The
reduced density matrix at a final time,
$\mel{s_{N}^{+}}{\tilde{\rho}\left(N\Delta t\right)}{s_{N}^{-}} = \tilde{\rho}\left(s_{N}^{\pm}, N\Delta t\right)$,
can be obtained using a discretized path integral representation as follows:
\begin{align}
  \tilde{\rho}\left(s_{N}^{\pm}, N\Delta t\right)  & = \Tr_{\text{b}}\mel{s_{N}^{+}}{\hat{U}(t) \tilde{\rho}(0)\otimes\rho_{\text{b}}(\beta) \hat{U}^{\dag}(t)}{s_{N}^{-}}\nonumber \\
                                                   & =\sum_{s_{0}^{\pm}} \tilde{\rho}\left(s_{0}^{\pm}, 0\right)G\left(s_{0}^{\pm},s_{N}^{\pm}, N\Delta t\right)\label{eq:RDM}      \\
  G\left(s_{0}^{\pm},s_{N}^{\pm}, N\Delta t\right) & = \sum_{s_{1}^{\pm}}\cdots\sum_{s_{N-1}^{\pm}} K_{1}\left(s_{0}^{\pm},s_{1}^{\pm}, \Delta t\right)\nonumber                    \\
                                                   & \times K_{2}\left(s_{1}^{\pm},s_{2}^{\pm}, \Delta t\right)\nonumber                                                            \\
                                                   & \cdots K_{N}\left(s_{N-1}^{\pm},s_{N}^{\pm}, \Delta t\right) F\left[\left\{s_{j}^{\pm}\right\}\right]\label{eq:green_func}
\end{align}
where $\Delta t$ is the quantum time-step that was used to discretize the path
integral. The augmented propagator for the reduced system, or the Green's
function, connecting state $s_{0}$ to $s_{k}$ in $k$ time steps is
$G(s_{0}^{\pm}, s_{k}^{\pm}, k\Delta t)$ and $\hat{U}$ is the full system-bath
propagator. The Feynman-Vernon influence functional~\cite{Feynman1963},
$F\left[\left\{s_{j}^{\pm}\right\}\right]$, encodes the interaction of the
system with the bath and is given as
\begin{align}
  F\left[\left\{s_{j}^{\pm}\right\}\right] & = \exp\left(-\frac{1}{\hbar}\sum_{0\le k\le N}\Delta s_{k}\sum_{0\le k'\le k}\left(\Re\left(\eta_{kk'}\right)\Delta s_{k'}\right.\right.\nonumber \\
                                           & \left.\vphantom{\sum_{0\le k\le N}}\left.+ 2i\Im\left(\eta_{kk'}\right)\bar{s}_{k'}\right)\right)\label{eq:influence_functional}
\end{align}
where $\eta_{kk'}$ are the discretized
$\eta$-coefficients~\cite{Makri1995_1,Makri1995_2},
$\Delta s_{k} = s^{+}_{k} - s^{-}_{k}$ and
$\bar{s}_{k} = \frac{1}{2}\left(s^{+}_{k}+s^{-}_{k}\right)$.

% Generally, \textcolor{blue}{the initial state of the system is known, and
%   therefore formulation of methods in terms of Eq.~\ref{eq:RDM} is most
%   convenient. Consequently, iterative QuAPI} directly calculates the reduced
% density matrix corresponding to an arbitrary initial state.
% However, there are certain advantages to calculating the augmented propagator using
% Eq.~\ref{eq:green_func}. Apart from being independent of the initial state of
% the system, the augmented propagator is also important for evaluation of the memory
% kernel associated with the Nakajima-Zwanzig~\cite{Nakajima1958,Zwanzig1960}
% generalized quantum master equation (GQME). Recently there has been a lot of
% interest in using
% approximate~\cite{Shi2003c,Montoya-Castillo2016,Kelly2016,Kidon2018} and
% numerically exact kernels for solving GQME~\cite{Chatterjee2019}.
% \textcolor{blue}{The small matrix decomposition of path integrals
%   (SMatPI)~\cite{Makri2020,Makri2020a} is defined in terms of this Green's
%   function.} Therefore, we formulate TNPI in terms of solving for the Green's
% function.

First, we define a couple terms that we use to rephrase the problem and build
the TNPI algorithm. Consider the bare and full variants of the path amplitude
tensor defined respectively as:
\begin{align}
  P^{(0)}_{s_{0}^{\pm}\cdots s_{N}^{\pm}} & =  K_{1}\left(s_{0}^{\pm},s_{1}^{\pm}, \Delta t\right) K_{2}\left(s_{1}^{\pm},s_{2}^{\pm}, \Delta t\right)\nonumber \\
                                          & \cdots K_{N}\left(s_{N-1}^{\pm},s_{N}^{\pm}, \Delta t\right)\label{eq:bare_path}                                    \\
  P_{s_{0}^{\pm}\cdots s_{N}^{\pm}}       & = F\left[\left\{s_{j}^{\pm}\right\}\right] P^{(0)}_{s_{0}^{\pm}\cdots s_{N}^{\pm}}.\label{eq:path_MPS}
\end{align}
The quantities $P^{(0)}$ and $P$ are tensors of rank $N$ with
$O\left(D^{2N}\right)$ coefficients. The bare PA tensor, $P^{(0)}$, contains the
full information about the bare system propagation. Information regarding any
time-dependence of the system Hamiltonian, $H_{0}$ is also included here. The
augmented propagator is obtained from the full PA tensor by contracting over all
but the terminal indices.
\begin{align}
  G\left(s_{0}^{\pm},s_{N}^{\pm}, N\Delta t\right) & = \sum_{s_{1}^{\pm}}\cdots\sum_{s_{N-1}^{\pm}} P_{s_{0}^{\pm}\cdots s_{N}^{\pm}}
\end{align}
% It is important to note here a basic difference between the structure of the
% augmented reduced density tensor (ADT)~\cite{Makri1995_1, Makri1995_2} and the
% PA that arises because of the incorporation of the initial condition
% $\tilde{\rho}\left(s_{0}^{\pm}, 0\right)$ into the ADT. This does not allow the
% calculation of the augmented propagator as independent from the initial condition.
% Therefore, the PA would play a fundamental role in the development of the
% method.
Since the number of values required to describe the PA tensor grows
exponentially with the number of time steps, a direct evaluation of the
augmented propagator is only possible if the number of time steps is small. The
goal of using tensor networks is to ``factorize'' these large tensors into
products of many smaller ones; and thus, reduce the cost of evaluation and
storage. The optimum structure of the tensor network involved is guided strongly
by the exact nature of the tensor under consideration.

It is well known that the non-Markovian memory length of a quantum system
interacting with a condensed phase is not infinitely long. Though these
non-local interactions do not have a finite support, they decay asymptotically.
The relative short-ranged nature of the interaction between time-points imply
that the PA tensor should be well described as a sequence of tensor products,
where each tensor corresponds to a different time-point. The resulting
factorization gives
\begin{align}
  P_{s_{0}^{\pm}\cdots s_{N}^{\pm}} & = \sum_{\left\{\alpha_{j}\right\}} M^{s_{0}^{\pm}}_{\alpha_{0}}M^{s_{1}^{\pm}}_{\alpha_{0}, \alpha_{1}}\cdots M^{s_{N-1}^{\pm}}_{\alpha_{N-2}, \alpha_{N-1}}M_{\alpha_{N-1}}^{s_{N}^{\pm}}\label{eq:PA_MPS}
\end{align}
which is the MPS representation of the PA tensor. In this representation, each
tensor in the product is considered to be on a different site. The indices that
are used in the superscript are the so-called ``site indices'' and they
correspond to the forward-backward state of the system at the different time
points. The indices in the subscript, $\left\{\alpha_j\right\}$, are the
so-called ``bond indices,'' the dimensions of which are closely related to the
length of the non-Markovian memory. Two relevant quantities that govern the
understanding of methods based on matrix product structures are the maximum bond
dimension, $m = \max(\dim(\alpha_j))$, and the average or typical bond
dimension, $\tfrac{1}{N}\sum_j \dim(\alpha_j)$. In principle this factorization
is exact; however, for an arbitrary PA tensor, the maximum bond dimension is
$O\left(D^{N}\right)$. In practice, the bond dimensions are truncated and
treated as convergence parameters. As we can see from Eq.~\ref{eq:PA_MPS},
truncating the bonds do not change the number of paths; therefore, all paths are
always considered. We will show in Sec.~\ref{sec:result}, that the actual bond
dimensions needed in a simulation are quite small.

\begin{figure}[h]
  \centering
  \includegraphics[scale=0.3]{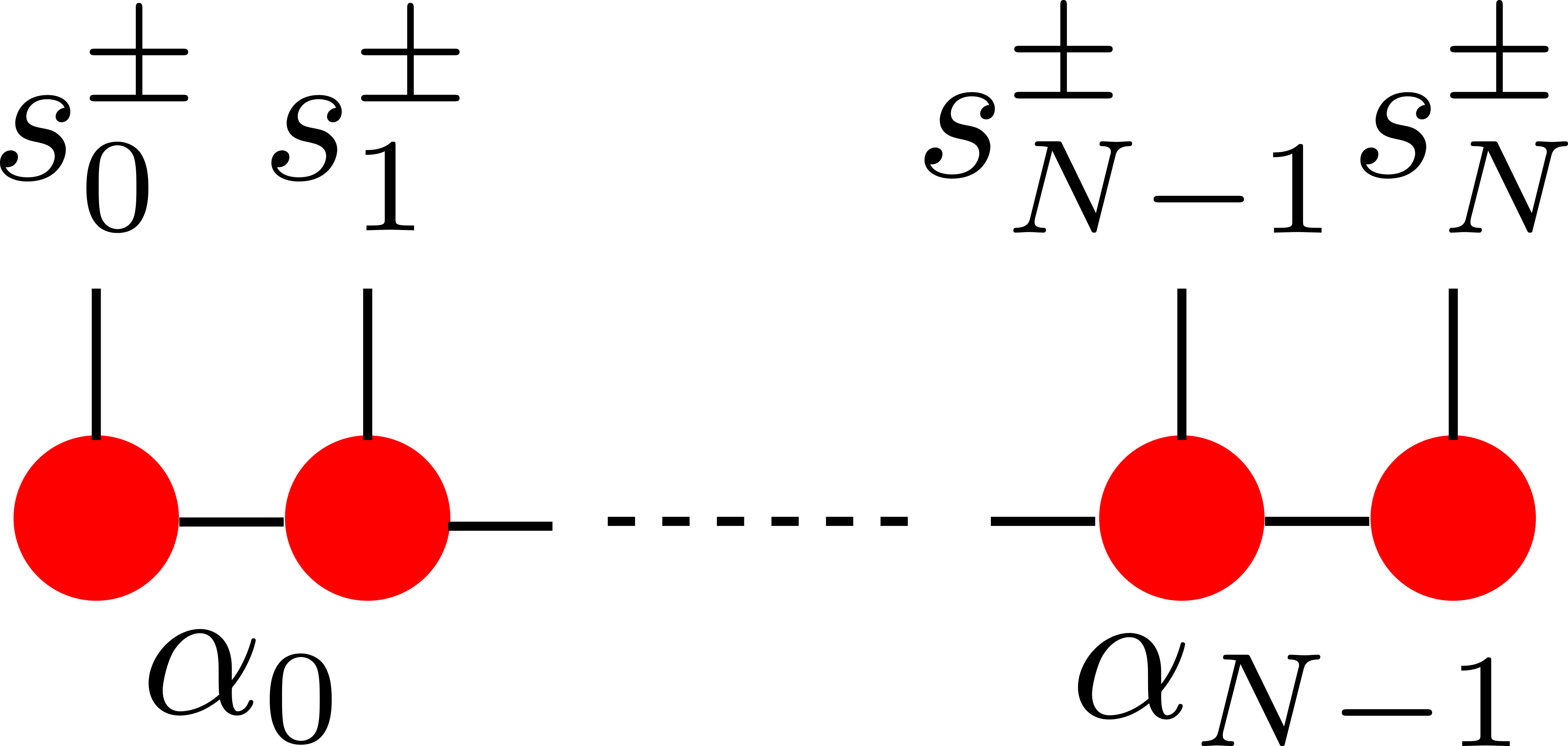}
  \caption{Schematic diagram for bare PA MPS. Red nodes denote the matrices corresponding to the bare system propagator's SVD.}
  \label{fig:barefig}
\end{figure}

The presence of the influence functional in the description of the full PA
tensor (Eq.~\eqref{eq:path_MPS}) prevents us from forming simple expressions for
the tensors in Eq.~\ref{eq:PA_MPS}; however, as we will now show, this is not
the case for the bare PA tensor. Consider the singular value decomposition (SVD)
of the forward-backward propagator:
\begin{align}
  K_{j}\left(s_{j-1}^{\pm},s_{j}^{\pm},\Delta t\right) & = \sum_{\alpha_{j-1}} U^{s_{j-1}^{\pm}} _{\alpha_{j-1}} \times R_{\alpha_{j-1}}^{s_{j}^{\pm}}\label{eq:SVDa} \\
  R_{\alpha_{j-1}}^{s_{j}^{\pm}}                       & = \sum_{\beta}  S_{\alpha_{j-1}, \beta}   \times V_{\beta}^{s_{j}^{\pm}\dag}\label{eq:SVDb}
\end{align}
where $S$ is the diagonal matrix of the nonzero singular values, $U$ and $V$
are the matrices of the left and right singular vectors respectively.
Substituting this into Eq.~\ref{eq:bare_path}, we get the required expressions
for the tensors appearing in Eq.~\ref{eq:PA_MPS}. Thus, for a bare PA MPS:
\begin{align}
  M_{\alpha_{0}}^{s_{0}^{\pm}}               & = U^{s_{0}^{\pm}} _{\alpha_{0}},                                         \\
  M^{s_{N}^{\pm}}_{\alpha_{N-1}}             & = R_{\alpha_{N-1}}^{s_{N}^{\pm}},                                        \\
  M_{\alpha_{j-1}, \alpha_{j}}^{s_{j}^{\pm}} & = R_{\alpha_{j-1}}^{s_{j}^{\pm}} U_{\alpha_{j}}^{s_{j}^{\pm}}\quad0<j<N.
\end{align}
It is worth noting that, in this case, the factorization is exact, and the bond
dimensions are independent of the number of time-steps, $N$. This is a direct
consequence of the Markovian nature of the system dynamics. We schematically
depict the bare PA as an MPS with red circular nodes as in
Fig.~\ref{fig:barefig}.

Now that we have described how to deal with the bare system, let us consider the
influence functional. From Eq.~\ref{eq:path_MPS}, we see that to obtain the full
PA MPS, the influence functional must be applied to the bare PA MPS. In the TN
framework, a MPO acts on a MPS to produce another MPS; therefore, we want to
construct an influence functional MPO, which, when applied to the bare PA MPS,
gives the PA MPS for the full system-bath problem. The full influence functional
can be rewritten as a product of terms coming from the interactions with
different end-points
\begin{align}
  F\left[\left\{s_{j}^{\pm}\right\}\right]     & = \prod_{0\le k\le N}F_{k}\left[\left\{s_{j}^{\pm}\right\}\right]                                                                 \\
  F_{k}\left[\left\{s_{j}^{\pm}\right\}\right] & = \exp\left(-\frac{1}{\hbar}\Delta s_{k} \sum_{0\le k'\le k} \left(\Re\left(\eta_{kk'}\right)\Delta s_{k'}\right.\right.\nonumber \\
                                               & \left.\phantom{\sum_{0\le k'\le k}}\left.+ 2i\Im\left(\eta_{kk'}\right)\bar{s}_{k'}\right)\right).\label{eq:if_last_pt}
\end{align}

The influence functional corresponding to a particular end time point,
$F_{k}\left[\left\{s_{j}^{\pm}\right\}\right]$ given by Eq.~\ref{eq:if_last_pt},
can be expressed as an MPO. We will construct this MPO in two steps by grouping
the forward-backward states by the unique values of $\Delta s$. First, consider the
trivial case when $\Delta s_{k} = 0$, and note that for all these states,
$F_{k} = 1$, or as an operator,
$F_{k} = \mathbb{I}^{s_{0}^{\pm}}\otimes\mathbb{I}^{s_{1}^{\pm}}\cdots \mathbb{I}^{s_{k-1}^{\pm}}\otimes\mathcal{P}^{s_{k}^{\pm}}_{0}$
where $\mathbb{I}^{s_{j}^{\pm}}$ is the identity operator in the space of
$\left\{s^{\pm}_{j}\right\}$, and $\mathcal{P}^{s_{k}^{\pm}}_{0}$ is the
projection operator on to the space where $\Delta s_{k} = 0$.

Next, we consider a non-zero value of $\Delta s_{k}$. For all these states, the
operators on sites, $k'<k$, are given as
$e^{-\tfrac{1}{\hbar}\Delta s_{k}\left(\Re\left(\eta_{kk'}\right)\mathbb{D}_{k'} + 2i\Im\left(\eta_{kk'}\right)\mathbb{S}_{k'}\right)}$,
and the operator on site $k'=k$ as
$\mathcal{P}^{s_{k}^{\pm}}_{\Delta s_{k}}e^{-\tfrac{1}{\hbar}\Delta s_{k}\left(\Re\left(\eta_{kk}\right)\mathbb{D}_{k} + 2i\Im\left(\eta_{kk}\right)\mathbb{S}_{k}\right)}$.
Here $\mathbb{D}_{k}$ and $\mathbb{S}_{k}$ are the matrices that return the
difference and average position of the system corresponding to the
forward-backward state. Both the $\mathbb{D}$ and $\mathbb{S}$ operators are
diagonal in the ${s^{\pm}}$ basis. For this particular value of $\Delta s_{k}$, the
resulting $F_{k}$, becomes direct product of these single site operators.
Therefore, $F_{k}$ can be represented as a sum of direct product of operators,
where each term in the sum corresponds to a different value of $\Delta s_{k}$. Hence,
we can define $F_{k}$ as a MPO, $\mathbb{F}_{k}$, as follows:
\begin{align}
  \mathbb{F}_{k} & = \sum_{\left\{\beta_{j}\right\}} W^{s_{0}^{\pm}, s_{0}^{'\pm}}_{\beta_{0}}(\eta_{k0})\cdots W^{s_{k'}^{\pm}, s_{k'}^{'\pm}}_{\beta_{k'-1}, \beta_{k'}} (\eta_{kk'})\nonumber \\
                 & \times W^{s_{k'+1}^{\pm}, s_{k'+1}^{'\pm}}_{\beta_{k'}, \beta_{k'+1}} (\eta_{k(k'+1)})\cdots W^{s_{k}^{\pm}, s_{k}^{'\pm}}_{\beta_{k-1}} (\eta_{kk})
\end{align}
where $\left\{\beta_{j}\right\}$ are the bond indices. In this case, all the bond
dimensions of this MPO are equal to the number of unique values of $\Delta s$.
We define $f(\beta)$ as the function that when given a value of the bond index $\beta$, returns
the corresponding value of $\Delta s$. Furthermore, we define
$\mathcal{P}^{s_{k}^{\pm}}_{f(\beta)}$ to be the projection operator on to the space
where $\Delta s_{k} = f(\beta)$. With this notation, now, we can define the
various tensors in our influence functional MPO.
\begin{widetext}
  \begin{align}
    W^{s_{0}^{\pm}, s_{0}^{'\pm}}_{\beta_{0}}(\eta_{k0})                    & =
    \delta_{s_{0}^{\pm}, s_{0}^{'\pm}}
    \exp\left(-\tfrac{1}{\hbar}f(\beta_{0}) (\Re(\eta_{k0}) \Delta s_{0} + 2i\Im(\eta_{k0}) \bar{s}_{0})\right)\label{eq:IFMPO_0}         \\
    W^{s_{k'}^{\pm}, s_{k'}^{'\pm}}_{\beta_{k'-1}, \beta_{k'}} (\eta_{kk'}) & =
    \delta_{s_{k'}^{\pm}, s_{k'}^{'\pm}}\delta_{\beta_{k'-1}, \beta_{k'}}
    \exp\left(-\tfrac{1}{\hbar}f(\beta_{k'-1}) (\Re(\eta_{kk'}) \Delta s_{k'} + 2i\Im(\eta_{kk'}) \bar{s}_{k'})\right)\label{eq:IFMPO_kp} \\
    W^{s_{k}^{\pm}, s_{k}^{'\pm}}_{\beta_{k-1}} (\eta_{kk})                 & =
    \delta_{s_{k}^{\pm}, s_{k}^{'\pm}} \mathcal{P}^{s_{k}^{\pm}}_{f(\beta_{k-1})}
    \exp\left(-\tfrac{1}{\hbar}\Delta s_{k} (\Re(\eta_{kk}) \Delta s_{k} + 2i\Im(\eta_{kk}) \bar{s}_{k})\right)\label{eq:IFMPO_k}
  \end{align}
\end{widetext}
Because $\mathbb{F}_{k}$ is an MPO representation of the influence functional
terms, it is invariant to any external time-dependence that the system
Hamiltonian might have. This analytic form of the IF MPO encodes the invariance
of the influence functional with respect to the average forward-backward system
states, $\bar{s}_k$, present in the form of Eq.~\ref{eq:influence_functional}.
This symmetry is also used by the blip-summed path
integral~\cite{Makri2014,Makri2014a} (BSPI) method as its starting point.
General algorithms have recently been developed to numerically generate the full
IF MPO, $\mathbb{F} = \prod_{k}\mathbb{F}_{k}$, for a larger class of
solvents~\cite{Ye2021,Cygorek2021} instead of the
$\left\{\mathbb{F}_{k}\right\}$ used here. While it might, in principle, be
possible to compress $\mathbb{F}_k$~\cite{Chan2016a,Parker2020}, we do not
foresee a great computational benefit. The analytical expressions already
utilize the inherent structure to construct an optimally compressed
representation. Additionally, it has been shown that filtration strategies
involving QuAPI influence functional generally work best when applied to the
product of the bare propagator and the
IF~\cite{Sim1997a,Makri2014,Makri2014a,Richter2017}.

\begin{figure}[h]
  \centering \subfloat[Calculation of $P^{(1)}$ cf. Eq.~\ref{eq:FullPathIter_0}.
    Red circles: SVD of forward-backward propagator $K$. MPS with black diamonds
    is the full PA MPS for two
    steps.]{\makebox[2.2\width][c]{\includegraphics[scale=0.3]{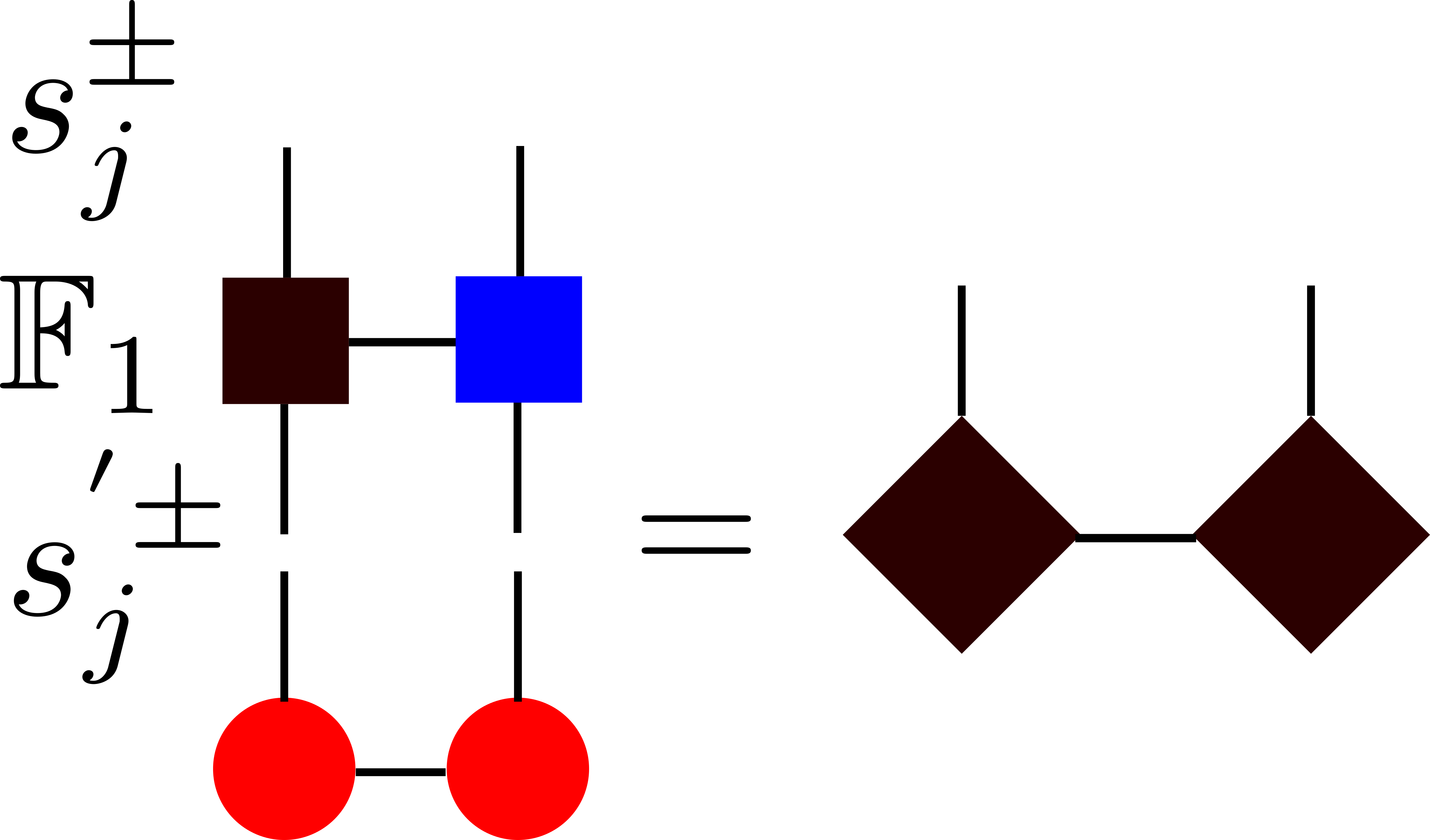}}}
  
  \subfloat[Calculation of $P^{(2)}$ cf. Eq.~\ref{eq:FullPathIter_k} starting
    from $P^{(1)}$ obtained in Fig.~\ref{fig:ifmposchema}~(a). Combination of dark
    red diamond and red circle comes from product of $P^{(1)}$ and $K$. MPS with
    black diamonds represents the full PA MPS for three
    steps.]{\makebox[1.5\width][c]{\includegraphics[scale=0.3]{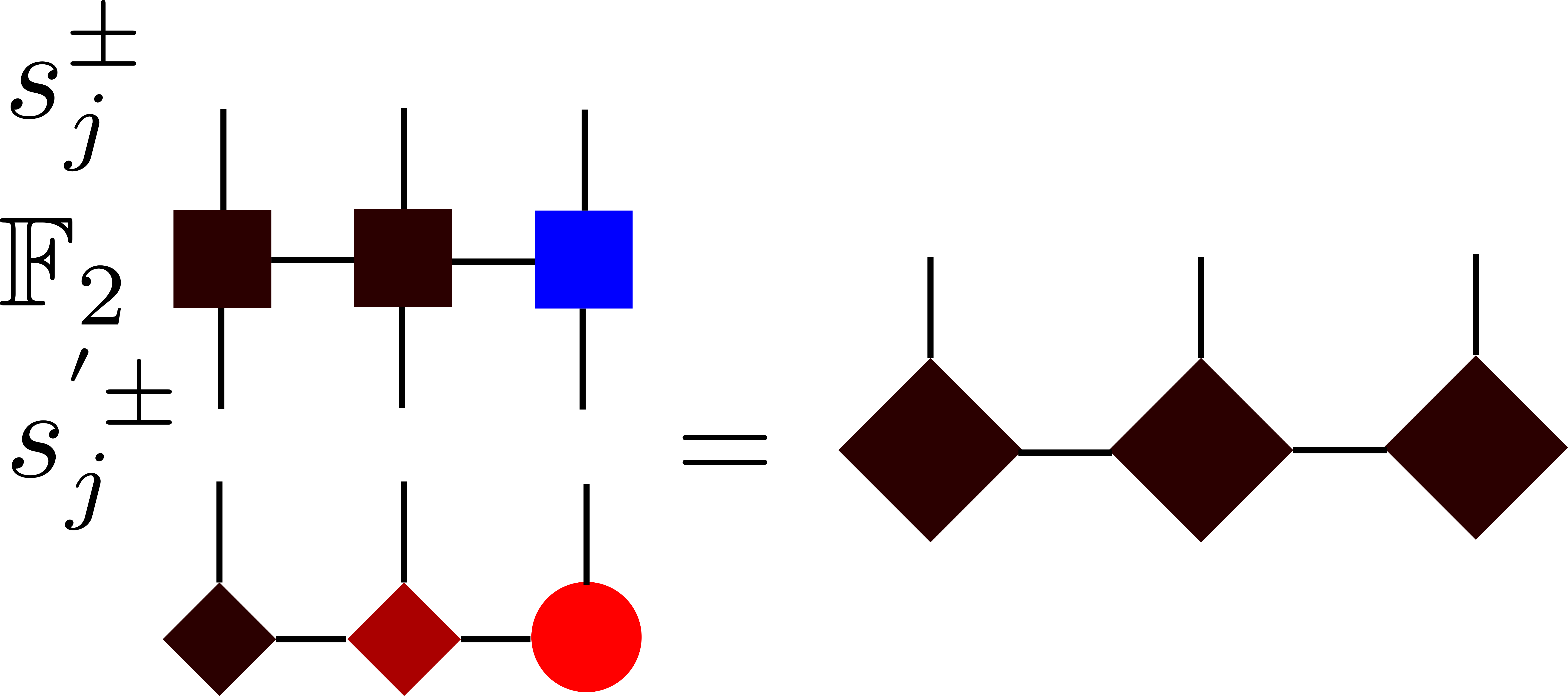}}}
  \caption{Schematic diagrams for application of IF MPO.}
  \label{fig:ifmposchema}
\end{figure}

The MPS representation of the full PA tensor can be obtained by iteratively
applying the $\mathbb{F}_{k}$ MPOs to the bare PA MPS.
\begin{align}
  P^{(1)}_{s_{0}^{\pm},s_{1}^{\pm}}       & = \sum_{\left\{\alpha_{j}^{'}\right\}}\mathbb{F}_{1} P^{(0)}_{s_{0}^{'\pm} ,s_{1}^{'\pm}}\label{eq:FullPathIter_0}                                                                      \\
  P^{(k)}_{s_{0}^{\pm}\cdots s_{k}^{\pm}} & = \sum_{\left\{\alpha_{j}^{'}\right\}}\mathbb{F}_{k} \left(P^{(k-1)}_{s_{0}^{'\pm}\cdots s_{k-1}^{'\pm}} K^{s_{k-1}^{'\pm},s_{k}^{'\pm}}_{\alpha_{k-1}}\right)\label{eq:FullPathIter_k} \\
  P_{s_{0}^{\pm}\cdots s_{N}^{\pm}}       & = \sum_{\left\{\alpha_{j}^{'}\right\}}\mathbb{F}_{N} \left(P^{(N-1)}_{s_{0}^{'\pm}\cdots s_{N-1}^{'\pm}} K^{s_{N-1}^{'\pm},s_{N}^{'\pm}}_{\alpha_{N-1}}\right)
\end{align}
where $\left\{\alpha_{j}^{'}\right\}$ are the new bonds indices produced by the
application of the MPO to the MPS. In the above expressions,
$P^{(k)}_{s_{0}^{\pm}\cdots s_{k}^{\pm}}$ represents the PA MPS after the
$k$\textsuperscript{th} step of the propagation, and the tensor
$K^{s_{k-1}^{\pm},s_{k}^{\pm}}_{\alpha_{k-1}}=U^{s_{k-1}^{\pm}}
  _{\alpha_{k-1}}R^{s_{k}^{\pm}}_{\alpha_{k-1}}$ and is associated with the SVD
decomposition of the $k$\textsuperscript{th} forward-backward system propagator,
$K_{k}\left(s_{k-1}^{\pm},s_{k}^{\pm},\Delta t\right)$ (see
Eqs.~\ref{eq:FB_Prop},~\ref{eq:bare_path},~\ref{eq:SVDa} and~\ref{eq:SVDb}). The
first couple of steps of the application of the IF MPO are schematically
portrayed in Fig.~\ref{fig:ifmposchema}. We would like to note that this clean
separation of the bare path amplitude being the MPS and the influence functional
being the MPO is not achieved in the time-evolving matrix product operator
method (TEMPO)~\cite{Strathearn2018}.

As we discussed earlier, the non-Markovian memory in a condensed phase process
decays to zero, thus we can construct an accurate MPS representation of the PA
tensor using fairly small values for the bond dimensions. However, the bond
dimensions can be further reduced by explicitly truncating the memory. To that
end, we define $L\Delta t$ to be the maximum distance, in time, between which
the influence functional can couple two forward-backward system points. We see
from Eq.~\ref{eq:green_func} that any system points that exceed this difference
can be summed independently of each other. This leads us to define the partially
contracted PA MPS, as well as the partially contracted IF MPO:
\begin{align}
   & \widetilde{P}_{s_{0}^{\pm},s_{N-L}^{\pm}\cdots s_{N}^{\pm}} =
  \sum_{s_{1}^{\pm}}\cdots \sum_{s_{N-L-1}^{\pm}} P_{s_{0}^{\pm}\cdots s_{N}^{\pm}},\quad L<N   \\
   & \widetilde{\mathbb{F}}_{k} =
  \sum_{\left\{\beta_{j}\right\}} W^{s_{0}^{\pm}, s_{0}^{'\pm}}_{\beta_{0}}(0)
  W^{s_{k-L}^{\pm}, s_{k-L}^{'\pm}}_{\beta_{0}, \beta_{k-L}} (\eta_{k(k-L)})\nonumber           \\
   & \quad \cdots W^{s_{k'}^{\pm}, s_{k'}^{'\pm}}_{\beta_{k'-1}, \beta_{k'}} (\eta_{kk'})\cdots
  W^{s_{k}^{\pm}, s_{k}^{'\pm}}_{\beta_{k-1}} (\eta_{kk}),\quad L<k
\end{align}
where the $W$ tensors are the same as those defined earlier (see
Eqs.~\ref{eq:IFMPO_0},~\ref{eq:IFMPO_kp} and~\ref{eq:IFMPO_k}).
% \textcolor{blue}{It can be seen that we have not contracted out the
%   $s_{0}^{\pm}$ index, unlike what is done in iteration schemes for
%   QuAPI~\cite{Makri1995_1,Makri1995_2} and related
%   methods~\cite{Makri2014,Makri2014a}, including TEMPO~\cite{Strathearn2018}.
%   This enables us to generate the augmented propagator independent of the initial
%   state of the system.}

We can now redefine the iterative procedure described above for calculating the
full PA MPS, in a manner consistent with an explicit truncation of the memory.
When $k\le L$, the procedure follows Eqs.~\ref{eq:FullPathIter_0}
and~\ref{eq:FullPathIter_k} exactly. This is called the ``full memory'' regime.
When $k>L$, which is often referred to as the ``iteration'' regime, we have
\begin{align}
  \widetilde{P}^{(L+1)}_{s_{0}^{\pm}\cdots s_{L+1}^{\pm}}           & = \sum_{\left\{\alpha_{j}^{'}\right\}} \widetilde{\mathbb{F}}_{L+1} P^{(L)}_{s_{0}^{'\pm} \cdots s_{L}^{'\pm}} K^{s_{L}^{'\pm},s_{L+1}^{'\pm}}_{\alpha_{L}}                               \\
  \widetilde{P}^{(k)}_{s_{0}^{\pm},s_{k-L}^{\pm}\cdots s_{k}^{\pm}} & = \sum_{\left\{\alpha_{j}^{'}\right\}} \widetilde{\mathbb{F}}_{k} \left(\sum_{s_{k-L-1}^{'\pm}}\widetilde{P}^{(k-1)}_{s_{0}^{'\pm},s_{k-L-1}^{'\pm}\cdots s_{k-1}^{'\pm}}\right.\nonumber \\
                                                                    & \left.\phantom{\sum_{s_{k-L-1}^{'\pm}}}\times K^{s_{k-1}^{'\pm},s_{k}^{'\pm}}_{\alpha_{k-1}}\right)                                                                                       \\
  \widetilde{P}^{(N)}_{s_{0}^{\pm},s_{N-L}^{\pm}\cdots s_{N}^{\pm}} & = \sum_{\left\{\alpha_{j}^{'}\right\}} \widetilde{\mathbb{F}}_{N} \left(\sum_{s_{N-L-1}^{'\pm}}\widetilde{P}^{(N-1)}_{s_{0}^{'\pm},s_{N-L-1}^{'\pm}\cdots s_{N-1}^{'\pm}}\right.\nonumber \\
                                                                    & \left.\phantom{\sum_{s_{N-L-1}^{'\pm}}}\times K^{s_{N-1}^{'\pm},s_{N}^{'\pm}}_{\alpha_{N-1}}\right)
\end{align}
Due to the aforementioned memory truncation, the number of system sites never
exceeds $L+2$. Lastly, the expression for the augmented propagator in terms of
the partially contracted PA MPS is given by
\begin{align}
  G\left(s_{0}^{\pm},s_{N}^{\pm}, N\Delta t\right) & = \sum_{s_{N-L}^{\pm}}\cdots\sum_{s_{N-1}^{\pm}} \widetilde{P}_{s_{0}^{\pm},s_{N-L}^{\pm}\cdots s_{N}^{\pm}}.
\end{align}
As it turns out, this iterative procedure is very similar to the one used by
iterative QuAPI. The key difference is that iterative QuAPI~\cite{Makri1995_1,
  Makri1995_2} and related methods~\cite{Makri2014a, Makri2014}, including ones
based on tensor networks~\cite{Jorgensen2019,Ye2021} such as
TEMPO~\cite{Strathearn2018}, sums over $s^{\pm}_{0}$ as soon as $k>L$, but here,
it is retained as part of the PA MPS and consequently the augmented propagator.
It is interesting that this calculation of the augmented propagator requires
negligible additional cost in the AP-TNPI framework. Apart from being important
for evaluation of the memory kernel associated with the
Nakajima-Zwanzig~\cite{Nakajima1958,Zwanzig1960} generalized quantum master
equation~\cite{Shi2003c,Montoya-Castillo2016,Kelly2016,Kidon2018,Chatterjee2019},
calculating the augmented propagator also allows the simulation to be
independent of initial state of the system. The small matrix decomposition of
path integrals (SMatPI)~\cite{Makri2020,Makri2020a}, which is designed to
minimize the storage requirements to roughly that of the reduced density matrix,
is also defined in terms of this augmented propagator.

There are various algorithms for applying an MPO to an MPS. Probably the two
most common methods are the direct application followed by an SVD-based
compression of the MPS and a variational application of the MPO to the MPS.
\citet{Paeckel2019a} have provided a detailed overview of these methods, the
computational effort involved and the common pitfalls. The scaling of the
variational MPO-MPS application for an MPS of length $l$ goes as $O(dlm^{3}w)$,
if the maximum MPS bond dimension is $m$, the MPO bond dimension is $w$, and the
site dimension is $d$~\cite{Paeckel2019a}. The site dimension is related to the
dimensionality of the quantum system as $d = D^{2}$. Though it is well-known
that the variational method can, depending on the initial ``guess'' MPS, get
stuck in metastable states, in our current explorations we have faced no such
issues. This might be because for a given time-step, the individual
$\mathbb{F}_{k}$ and $\widetilde{\mathbb{F}}_{k}$ do not significantly alter the
MPSs. It is interesting to note that this ``perturbative'' impact of the
influence functional MPO on the path amplitude MPS is a result of the particular
decomposition achieved by separating the bare propagator part completely from
the influence functional. However, we still checked all the calculations
reported against the direct SVD method as well.

In addition to the iterative algorithms we have outlined till now, AP-TNPI
allows a different approach that uses the augmented propagator between $t=0$ and
$t=L\Delta t$ to approximate the propagation when the Hamiltonian is
time-independent:
\begin{align}
  \tilde{\rho}\left(s_{k}^{\pm}, k\Delta t\right) & = \sum_{s_{0}^{\pm}}G\left(s_{0}^{\pm},s_{k}^{\pm}, k\Delta t\right) \tilde{\rho}\left(s_{0}^{\pm}, 0 \right) ,\quad \text{when } k \le L\label{eq:MTNPIa} \\
  \tilde{\rho}\left(s_{k}^{\pm}, k\Delta t\right) & \approx \sum_{s_{k-L}^{\pm}}G\left(s_{0}^{\pm},s_{L}^{\pm}, L\Delta t\right)\nonumber                                                                      \\
                                                  & \times\tilde{\rho}\left(s_{k-L}^{\pm}, (k-L) \Delta t \right) \delta_{s_{k-L}^{\pm}, s_{0}^{\pm}}, \quad \text{otherwise.}\label{eq:MTNPIb}
\end{align}
We call this scheme the Markovian iteration TNPI scheme (MTNPI). While it is
exact within memory, MTNPI misses out on some interactions present in the
canonical iteration scheme, and consequently converges slower with respect to
$L$. However, if this increased memory can be spanned, the iteration is just a
sequence of matrix-vector multiplications, and hence very fast. Even when memory
is not converged, this method results in computationally cheap dynamics that is
quite physically insightful. This computational advantage becomes especially
important when the total time-scale of the dynamics far exceeds the length of
the non-Markovian memory.

Please note that while for simplicity the entire development in this section has
been done in terms of a bi-linearly coupled bath that interacts with the
``position'' operator of the system, this is not a necessary restriction. The
formalism of Feynman-Vernon influence functional is general, and consequently
AP-TNPI can also be extended to deal with other couplings. The more general
expressions presented in Refs.~\cite{Strumpfer2009,Nalbach2010} are summarized
in Appendix~\ref{app:if_local} for convenience.

\section{Results}\label{sec:result}

We will now apply AP-TNPI to a large variety of illustrative examples. This
section is organized as follows: first, we benchmark the method on two-level
systems against other path integral methods. We try to
understand the behavior of the bond dimensions and convergence in general. We
show that AP-TNPI can easily simulate dynamics which can only be treated with
more recent
techniques~\cite{Walters2016,Makri2014,Makri2014a,Makri2020,Makri2020a}. Then,
we apply AP-TNPI to two realistic applications: a charge transfer example and
excitonic dynamics in a dimer. For the charge transfer, using a combination of
rate theory and direct dynamics, we try to understand the essential physics,
whereas for the excitonic dynamics, we are able to converge the relevant
dynamics directly. Finally, we show examples of applying AP-TNPI to multilevel
systems.

Before continuing to the results, a brief mention of the convergence parameters
is in order. Apart from the standard time step ($\Delta t$) and memory length ($L$)
associated with iterative path integral methods, AP-TNPI has two additional
controllable parameters: the maximum allowable bond dimension ($M$), and the
truncation threshold ($\chi$). While the precise meaning of the truncation
threshold depends of the exact nature of the compression algorithm used,
here it can be thought of as a SVD-like truncation, where any
singular values smaller then the given threshold are omitted. More
specifically, the singular values, $\lambda_{n}$, are discarded such that
\begin{align}
  \frac{\sum_{n\in\text{discarded}}\lambda_{n}^{2}}{\sum_{n}\lambda_{n}^{2}}<\chi.
\end{align}
These four parameters are interrelated, therefore, the
convergence is ultimately a search procedure to find the combination that gives
the correct dynamics at the least cost. For simplicity, we generally have not
explicitly constrained the maximum bond dimension.

\subsection{Two-level systems}
Two-level systems (TLS) are some of the most commonly studied model systems in quantum
dynamics. They are relatively simple and yet support a surprisingly rich array
of dynamics. As the name implies, the system is described by a Hamiltonian with
2 states:
\begin{align}
  \hat{H}_{0} & = \sum_{j=1}^{2}\epsilon_{j}\dyad{\sigma_{j}} - \hbar\Omega\left(\dyad{\sigma_{1}}{\sigma_{2}} + \dyad{\sigma_{2}}{\sigma_{1}}\right),\label{eq:TLS}
\end{align}
where $\ket{\sigma_{1}}$ and $\ket{\sigma_{2}}$ are eigenstates of the system's position
operator, $\hat{s}$, i.e., $\hat{s}\ket{\sigma_{j}}=\sigma_{j}\ket{\sigma_{j}}$; throughout
this subsection, this pair of eigenvalues, $\{\sigma_{1},\sigma_{2}\}$, will be referred
to as the the system's coordinates. Additionally, the position dependent
energies are defined as $\epsilon_{1}=\epsilon$ and $\epsilon_{2}=-\epsilon$. With this model, the key
parameters are the asymmetry, $\epsilon$, and the tunneling splitting of $2\hbar\Omega$. All the
examples in this subsection use Eq.~\ref{eq:TLS} as the Hamiltonian for the
isolated system.

\subsubsection{Benchmark calculations}
For the benchmark calculations, we track the population decay of the system.
More specifically, we simulate $\mel{1}{\tilde{\rho}\left(t\right)}{1}$, where
$\mel{1}{\tilde{\rho}\left(0\right)}{1}=1$. The system is taken to be coupled
bi-linearly with a harmonic bath as described by Eq.~\ref{eq:Hamil}, with the
spectral density having the Ohmic form
\begin{align}
  J(\omega) = \frac{\pi}{2}\hbar\xi\omega\exp(-\frac{\omega}{\omega_{c}}).\label{eq:spectral}
\end{align}
Here, $\xi$ is the dimensionless Kondo parameter that characterizes the
system-bath coupling, and $\omega_{c}$ is the cutoff frequency of the bath. To keep
these calculations general, the parameters are given in terms of $\Omega$.
\begin{figure}[h]
  \centering
  \subfloat[Population dynamics]{\includegraphics{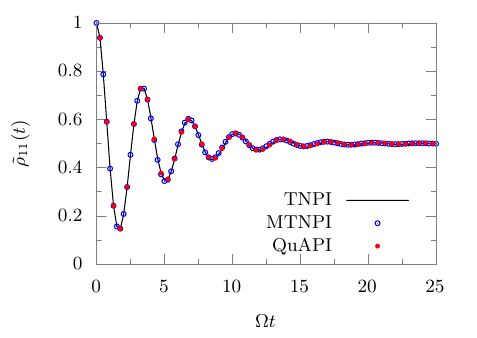}}
  
  \subfloat[Off-diagonal terms. Solid line, markers: $\Re(\tilde{\rho}_{01})$. Dashed
    line, hollow markers: $\Im(\tilde{\rho}_{01})$. Lines: TNPI. Red circles: iQuAPI.
    Blue triangles: MTNPI]{\includegraphics{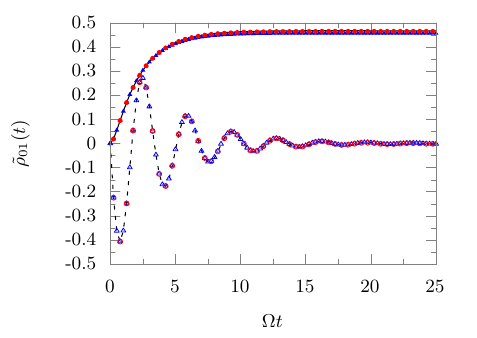}}
  \caption{Comparison with QuAPI for typical spin-boson parameters. Bath is
  characterized by $\hbar\Omega\beta=5$, $\omega_{c} = 7.5\Omega$, $\xi = 0.1$,
  $\chi=10^{-6}$.}\label{fig:spin-boson-comp}
\end{figure}

In Fig.~\ref{fig:spin-boson-comp}, we compare fully converged AP-TNPI results
with QuAPI for a symmetric ($\epsilon = 0$) system interacting weakly with a
cold, fast harmonic bath, as represented by $\hbar\Omega\beta=5$, $\omega_{c} =
  7.5\Omega$ and $\xi = 0.1$. The system's coordinates are $\{1,-1\}$. Because
there are no additional approximations, both AP-TNPI and QuAPI converge with
exactly the same $\Omega\Delta t = 0.25$ and $L = 8$. The same level of
compression was used to converge both the diagonal and the off-diagonal terms of
the RDM. We show the converged results for MTNPI done at $L=15$. Despite the
additional memory introduced by the simplified MTNPI propagation scheme, the
basic physics is still reproduced. For this parameter, using MTNPI we were able
to reproduce the population curves and the imaginary part of $\rho_{01}$
exactly. Though rather accurate, there is a very slight discrepancy in the MTNPI
estimation of the real part of $\rho_{01}$.

\begin{table*}[]
  \begin{tabular}{r|cccccc|r}
    L  & $10^{-4}$   & $10^{-6}$   & $10^{-8}$     & $10^{-10}$     & $10^{-12}$      & $0$               \\ \hline
    2  & 4, 4, 0.026 & 4, 4, 0.026 & 4, 4, 0.026   & 4, 4, 0.026    & 4, 4, 0.026     & 4, 4, 0.026       \\
    4  & 4, 4, 0.019 & 8, 6, 0.011 & 8, 6, 0.011   & 12, 7, 0.011   & 16, 8, 0.011    & 16, 9, 0.011      \\
    6  & 4, 4, 0.026 & 8, 6, 0.006 & 9, 6, 0.006   & 19, 10, 0.005  & 20, 11, 0.006   & 64, 25, 0.005     \\
    8  & 4, 4, 0.028 & 8, 7, 0.002 & 9, 7, 0.002   & 22, 13, 0.001  & 25, 14, 0.001   & 256, 76, 0.001    \\
    10 & 4, 4, 0.029 & 8, 7, 0.001 & 11, 9, 0.0004 & 21, 14, 0.0004 & 26, 17, 0.0001  & 1024, 248, 0.0001 \\
    12 & 4, 4, 0.030 & 8, 7, 0.001 & 11, 9, 0.0007 & 20, 15, 0.0001 & 27, 19, 0.00001 & 4096, 840, 0
  \end{tabular}
  \caption{Maximum and Average bond dimension at end point of simulation (rounded
    to the nearest integer) for different values of $\chi$ (on different columns)
    and the average RMS errors associated with all RDM elements for each simulation.
    The bond dimensions for the uncompressed MPS (at $\chi=0$) are also shown for
    comparison.}\label{tab:bonddim}
\end{table*}

% \begin{table}[]
%   \begin{tabular}{r|cccccc|r}
%     L  & $10^{-4}$  & $10^{-6}$ & $10^{-8}$ & $10^{-10}$ & $10^{-12}$  & $0$         & \# paths \\ \hline
%     2  & 4, 0.026   & 4, 0.026  & 4, 0.026  & 4, 0.026   & 4, 0.026    & 4, 0.026    & 64       \\
%     4  & 4, $0.019$ & 6, 0.011  & 6, 0.011  & 7, 0.011   & 8, 0.011    & 9, $0.011$  & 1024     \\
%     6  & 4, $0.026$ & 6, 0.006  & 6, 0.006  & 10, 0.005  & 11, 0.006   & 25, 0.005   & 16384    \\
%     8  & 4, $0.028$ & 7, 0.002  & 7, 0.002  & 13, 0.001  & 14, 0.001   & 76, 0.001   & 262144   \\
%     10 & 4, $0.029$ & 7, 0.001 & 9, 0.0004 & 14, 0.0004 & 17, 0.0001  & 248, 0.0001 & 4194304  \\
%     12 & 4, $0.030$ & 7, 0.001 & 9, 0.0007 & 15, 0.0001 & 19, 0.00001 & 840, 0      & 67108864
%  \end{tabular}
% \caption{Average bond dimension at end point of simulation (rounded to the
% nearest integer) for different values of $\chi$ (on different columns) and the
% average RMS errors associated with all RDM elements for each simulation. The
% number of system paths in a typical path integral calculation and the bond
% dimensions for the uncompressed MPS (at $\chi=0$) are also shown for
% comparison.}\label{tab:bonddim}
% \end{table}

\begin{figure}[h]
  \includegraphics{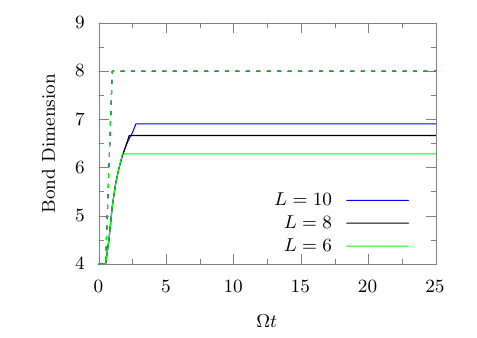}
  \caption{Growth of average (solid) and maximum (dashed) bond dimensions of PA
    MPS with time and $L$ for parameters shown in Fig.~\ref{fig:spin-boson-comp}
    calculated with $\chi = 10^{-6}$.}\label{fig:avgbonddim}
  % \caption{Growth of average bond dimension of PA MPS with time and $L$ for
  %   parameters shown in Fig.~\ref{fig:spin-boson-comp}. Black dashed line:
  %   full path. Gray solid line: $L = 20$. Gold solid line: $L = 17$. Magenta
  %   solid line: $L = 14$. Blue solid line: $L = 11$. Black solid line: $L =
  %   8$. Green solid line: $L = 5$. Red solid ine: $L=
  %   2$.}\label{fig:avgbonddim}
\end{figure}

Before showing more computationally intensive examples, let us analyze the
growth of the bond dimension with the memory length $L$ and the cutoff $\chi$
for the simple case of Fig.~\ref{fig:spin-boson-comp}. In
Table~\ref{tab:bonddim}, we show the maximum and average bond dimensions
corresponding to calculations with particular memory length, $L$, at the end of
the simulation time of $\Omega t = 25$, and the average root-mean-square (RMS)
deviation for all the RDM elements from the most accurate simulation done. Since
the cost of the MPO-MPS application scales as the cube of the average MPS bond
dimension~\cite{Paeckel2019a}, setting $\chi=0$ results in a calculation that is
even more costly than QuAPI. However, even with a minimal value of $\chi>0$,
there is a dramatic reduction of the bond dimensions. Additionally, we see that
when the memory length is far from convergence, it is difficult to judge the
convergence of $\chi$, and vice versa. (While it is interesting to study the RMS
errors for these benchmark simulations, it cannot serve as a sole guide to
convergence. This is particularly true, if the results aren't known beforehand.
When unconverged, the dynamics may still show unphysical behaviors despite the
calculated error being quite small.) The time-dependence of the average
bond-dimension for different values of $L$ near convergence is given in
Fig.~\ref{fig:avgbonddim}. Here, we see that during the iteration, the
bond-dimension remains practically constant; therefore, the computational cost
scales linearly with the number of steps. Notice that for this problem, the
converged cutoff is high enough that the maximum bond dimension for all three
memory lengths is the same.

\begin{figure}[h]
  \includegraphics{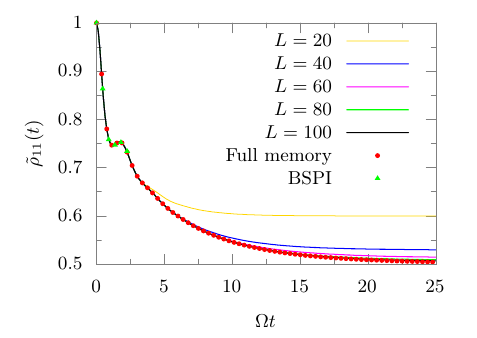}
  \caption{Simulation with very long non-Markovian memory: $\epsilon=0$, $\hbar\Omega\beta=1$,
    $\omega_{c} = \Omega$, $\xi = 2$. BSPI results are shown up to the point where they
    could be converged.}\label{fig:spin-boson-long-mem}
\end{figure}
\begin{figure}[h]
  \includegraphics{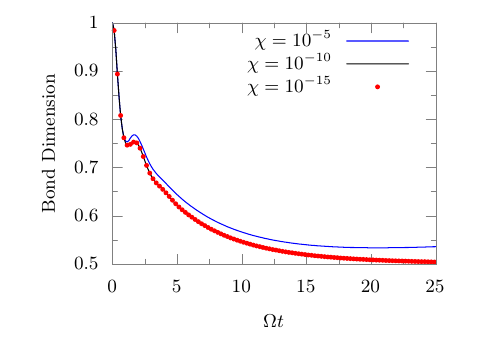}
  \caption{Dynamics for different values of $\chi$ when $M$ is left unconstrained.
    Parameters same as Fig.~\ref{fig:spin-boson-long-mem}. These are full memory
    calculations.}\label{fig:infmaxbond}
\end{figure}
In Fig.~\ref{fig:spin-boson-long-mem}, we consider a symmetric TLS ($\epsilon =
  0$) strongly coupled ($\xi = 2$) to a sluggish bath ($\omega_{c} =
  \Omega$)~\cite{Walters2015,Walters2016}. The system's coordinates are
$\{0,-2\}$, i.e., the bath is initially equilibrated to the populated state of
the system. Here, $\Omega\Delta t = 0.125$ and $\chi = 10^{-10}$. The strong
coupling and sluggish bath imply that the non-Markovian memory is quite long.
Generally, this is the regime where iterative QuAPI performs the worst. It is
interesting to note that with AP-TNPI, owing to the compressions enabled by the
MPS representation, we can converge the dynamics directly till equilibration of
the system around $\Omega t \approx 25$, reproducing results previously obtained
by SMatPI~\cite{makriSmallMatrixPath2021}. Convergence is reached at a memory
length of $L = 100$. In this particular case, AP-TNPI was remarkably successful
at compressing the PA MPS, which at the final time point of $\Omega t = 25$ has
an average bond dimension of $M = 35.96$. For comparison, we were unable to
converge the full dynamics using blip-summed path integral
(BSPI)~\cite{Makri2014,Makri2014a}. This is because though this is a case of a
strongly coupled slow bath, the conditions are not close enough to an incoherent
decay to enable convergence based on the filtration by the number of blips, i.e.
time points where $\Delta s \ne 0$.

We also simulate the full memory dynamics using various cutoffs in
Fig.~\ref{fig:infmaxbond}. Even at a fairly large cutoff of $\chi=10^{-5}$,
though the dynamics is not converged, the key features are preserved. Although
the maximum bond dimension, with $\chi=10^{-5}$, is never greater then eight,
the resulting error in dynamics is smaller than that produced from a memory
truncation of $L=20$. Given how the bond truncation works when applying a MPO to
a MPS, even with relatively large values of $\chi$, unitarity of propagation is
preserved and consequently, the trace is conserved.

\subsubsection{Realistic applications of spin-boson model}
The compressed representation enabled by AP-TNPI allows us to span longer memory
lengths. This makes it possible to use AP-TNPI to, at least qualitatively,
understand the physics of realistic systems. As the first example, consider the
charge transfer in the ferrocene-ferrocenium complex in hexane as a solvent.
This system was studied using QCPI by one of us~\cite{Walters2015a}. It was
shown that the harmonic bath results match the fully atomistic results really
well. We apply AP-TNPI to understand the dynamics qualitatively using only short
calculations that can be run on extremely modest computational resources.

\begin{figure}[h]
  \centering
  \includegraphics{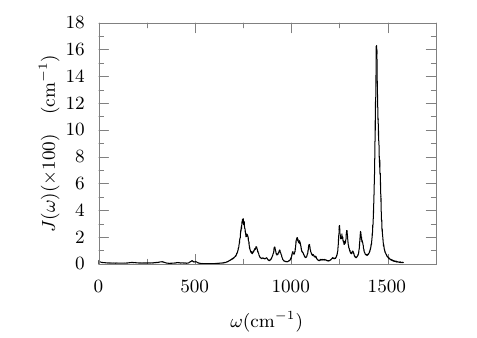}
  \caption{Spectral density for hexane interacting with ferrocene-ferrocenium.}
  \label{fig:hexanespect}
\end{figure}
The system is described by a symmetric TLS Hamiltonian, where $\hbar\Omega =
  \SI{32}{\per\centi\meter}$ and $\epsilon = 0$. As in Eq.~\ref{eq:Hamil}, the
system is coupled bi-linearly to the bath, which is initially equilibrated with
the donor state. The spectral density of the harmonic bath with a reorganization
energy of $\SI{36}{\milli\eV}$~\cite{Walters2015a} is shown in
Fig.~\ref{fig:hexanespect}. It is extremely challenging to converge the full
dynamics using the basic AP-TNPI framework outlined here because of long memory
times, however, we can quite easily converge the initial decay of the
population. It is the presence of classical trajectories in QCPI that allows for
accelerated convergence with respect to memory length and incorporation of
anharmonic effects. Such improvements can, in principle, be added into AP-TNPI,
and will be the subject of future research. Here we use the basic AP-TNPI
framework outlined in this paper.

\begin{figure}[h]
  \centering
  \includegraphics{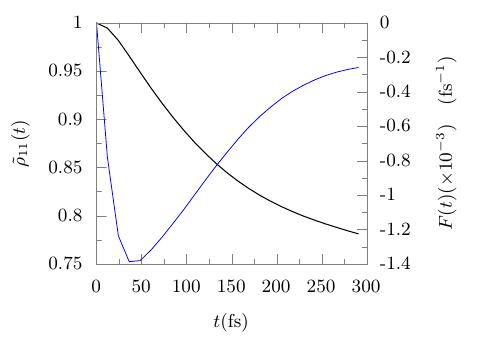}
  \caption{Population of donor state (black line, left axis) and flux of the
    donor state (blue line, right axis) as a function of time.}
  \label{fig:ffdyn}
\end{figure}

For reactions with long time-scales, rate theory can often be used to gain
valuable insights. Quantum mechanical rates are typically related to equilibrium
correlation functions involving the flux operator, $\hat{F} =
  \tfrac{i}{\hbar}\comm{\hat{H}_{0}}{\hat{h}}$ where $\hat{h} = \dyad{1}$ is the
projector on the donor state,
$\ket{1}$~\cite{Miller1975,Miller1983,Yamashita1985,Topaler1994,Bose2021a}.
Since the equilibrium of a system coupled to a solvent can often be challenging
to compute, it has also been shown that the same information can be obtained
from the non-equilibrium flux function~\cite{Bose2017}, which also happens to be
related to the instantaneous time-derivative of the donor population. In
Fig.~\ref{fig:ffdyn}, we show the short-time dynamics of both the donor
population, $\tilde\rho_{11}(t)$, and the donor flux, $F(t) =
  \tfrac{d\tilde\rho_{11}}{dt} = \Tr(\tilde\rho(t) \hat{F})$, as calculated by
AP-TNPI. The simulation was run at $T=\SI{300}{\kelvin}$. Convergence was
reached with a time-step of $\Delta t\approx \SI{12}{\femto\second}$. These full
memory simulations were done for 24 time steps and the truncation cutoff was
converged at $\chi = 10^{-16}$. In this case, we plot the dynamics corresponding
to a maximum bond dimension of $M = 1000$, though $M = 750$ yielded converged
results in the timescale shown. A laptop with a total RAM of 8\,GB was used for
the computation. With access to larger computer memory, it might well be
possible to converge the full dynamics directly with AP-TNPI. This real world
example is very similar to the sort of parameters that we considered in
Fig.~\ref{fig:spin-boson-long-mem}. The dynamics is a monotonic decay, but still
paths containing multiple blips states, i.e. states with $\Delta s \ne 0$, are
very important. This proves to be exceedingly challenging for BSPI as well.

The population has not decayed to equilibrium within the first
$\SI{300}{\femto\second}$. However, a lot of information about the dynamics can
be gleaned from a combination of the direct dynamics and the flux function (blue
curve). If rate theory is valid, and a slow exponential decay governs the
dynamics, then the flux function should quickly plateau at the rate after
transients. We see that the flux function in this case shows dramatic changes
implying that the transients are not completely over. Combined with the not
insignificant population decay, the dramatic changes in the flux function
demonstrates the non-exponential nature of the dynamics for this system
reproducing the conclusions of the initial exploration~\cite{Walters2015a}.
While exact details about the dynamics cannot be inferred unless either the
dynamics is simulated to equilibrium or the flux function is simulated to a
plateau, it is possible to make intelligent conclusions about the various
timescales. The deep drop in the flux function, representing a fast population
decay, is followed by a rapid increase, representing a significant slowdown of
the transfer dynamics. This large maximum flux out of the donor state, coupled
with the non-insignificant decay of population dynamics within the time of
simulation allows us to draw intelligent conclusions about the timescales. A
``dominant'' $\tfrac{1}{e}$--time of $\sim\SI{720}{\femto\second}$ can be
estimated from the maximum flux out of the donor state. Interestingly, this
value agrees with the reported $\tfrac{1}{e}$--time of $\SI{830}{\femto\second}$
calculated from the full QCPI dynamics. The flux curve also shows a significant
slowing down of the dynamics in a non-exponential manner. Consequently,
traditional equilibrium rate theory would give incorrect rates and predict much
longer time-scales. For extremely non-exponential processes like this particular
one, it becomes impossible to predict the long-time timescales from such short
time dynamics because the flux function does not plateau.

Though the method can be readily extended to larger systems, let us, now, turn
our attention to exciton transfer in a dimer. This can be considered the
simplest illustrative example of an exciton transfer. Restricting ourselves to
the single-exciton Frenkel subspace, we have a two-level description for the
system where $\ket{\sigma_{1}}$ corresponds to the first monomer being in the excited
state and the second in the ground state; and $\ket{\sigma_{2}}$ corresponds to the
second monomer being in the excited state and the first in the ground state. The
vibrational degrees of freedom of each of the monomers produce a dissipative
bath. The Hamiltonian of the bath and its interaction with the system is now
given by
\begin{align}
  \hat{H}_{\text{b}} & = \sum_{j,b} \frac{p_{jb}^{2}}{2m_{jb}} + \frac{1}{2}m_{jb}\omega_{jb}^{2}\left(x_{jb} - \frac{c_{jb}\hat{V}_{j}}{m_{jb}\omega_{jb}^{2}}\right)^{2}\label{eq:exiton_Hb}
\end{align}
where the $b$\textsuperscript{th} vibrational mode associated with the
$j$\textsuperscript{th} monomer has a frequency of $\omega_{jb}$ and coupling
constant of $c_{jb}$. Additionally, $\hat{V}_{j} = \dyad{\sigma_{j}}$ is the projector on to
state where the $j$\textsuperscript{th} monomer is excited. This coupling to the
system captures the shift of the Born-Oppenheimer surfaces of the localized
vibrations between the ground and the excited electronic states of the monomer.

\begin{figure}[h]
  \centering
  \includegraphics{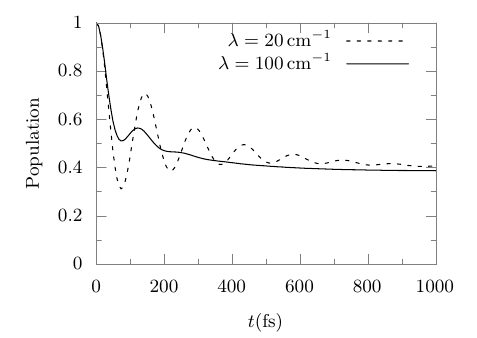}
  \caption[Excitonic dynamics]{Excitonic dynamics in a dimer
    $\left(\epsilon=\SI{50}{\per\centi\meter}\right.$,
      $\left.\hbar\Omega=-\SI{100}{\per\centi\meter}\right)$ with a Drude bath with
    $\gamma=\SI{53.08}{\per\centi\meter}$ at $T=\SI{300}{\kelvin}$.}
  \label{fig:exciton}
\end{figure}
Since the two monomers are identical molecules, the spectral densities for the
localized molecular vibrations are taken to be identical for both the baths.
While AP-TNPI is able to handle discrete rigid vibrations directly, first, we
follow \citet{Ishizaki2009} and assume that the bath is described by a Drude
spectral density,
\begin{align}
  J(\omega) & = 2\lambda\gamma\frac{\omega}{\omega^{2} + \gamma^{2}},
\end{align}
where $\lambda$ is the reorganization energy and $\gamma$ is the cutoff frequency.
Figure~\ref{fig:exciton} shows the dynamics of the population corresponding to
$\ket{\sigma_{1}}$ in the excitonic dimer defined by
$\hbar\Omega =- \SI{100}{\per\centi\meter}$, $\epsilon = \SI{50}{\per\centi\meter}$ and
$\gamma=\SI{53.08}{\per\centi\meter}$ at two different values of reorganization
energies \SIlist{20;100}{\per\centi\meter}. These simulations were performed at
room temperature, i.e., $T=\SI{300}{\kelvin}$. A time step of
$\Delta t = \SI{6.05}{\femto\second}$ was used with a memory length of $L = 80$ and a
truncation threshold of $\chi=10^{-10}$. The results are same as the ones reported
by Ishizaki and Fleming using HEOM.

\begin{figure}[h]
  \centering
  \includegraphics{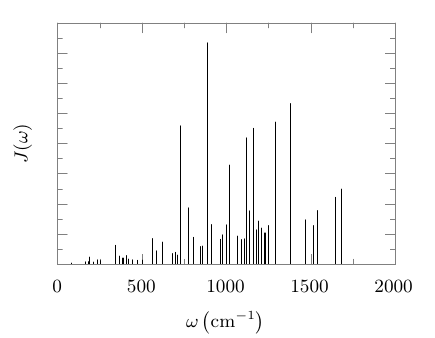}
  \caption{Vibrational spectral density for a bacteriochlorophyll molecule in
    arbitrary units with a reorganization energy of
    $\lambda = \SI{217.66}{\per\centi\meter}$.}
  \label{fig:spectral_density}
\end{figure}
Though, till now, we discussed the dynamics corresponding to the Drude spectral
density, chlorophyll, being a rigid monomer, has a much more structured,
discrete vibrational spectral density. The detailed impact of such structured
vibrations on the dynamics cannot be captured by using model spectral densities.
Figure~\ref{fig:spectral_density} shows the spectral density obtained from the
Huang-Rhys factors reported by~\citet{Ratsep2011} with a reorganization energy
$\lambda = \SI{217.66}{\per\centi\meter}$. This vibrational bath has already
been used to study the quantum dynamics of a bacteriochlorophyll dimer using
QCPI~\cite{Bose2020} and for longer chains and rings~\cite{Kundu2020a} using
modular path integral~\cite{Makri2018,Makri2018a,Kundu2019}.

\begin{figure}[h]
  \centering
  \includegraphics{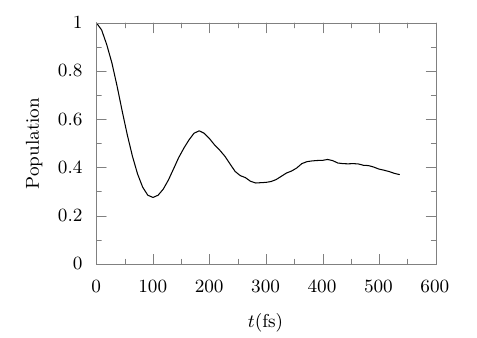}
  \caption{Excitonic dynamics of the excited monomer in a dimer with the rigid,
    structured bath given in Fig.~\ref{fig:spectral_density}. The system
    Hamiltonian is the same as used in Fig.~\ref{fig:exciton}.}
  \label{fig:structured_dynamics}
\end{figure}
\begin{figure}[h]
  \centering
  \includegraphics{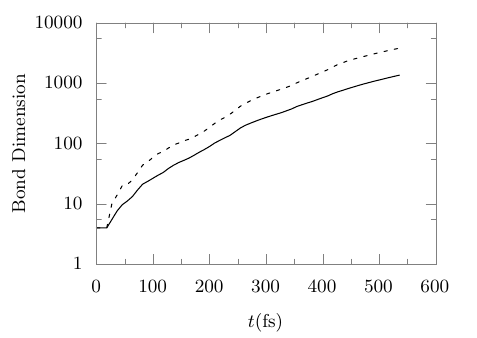}
  \caption{Growth of average (solid) and maximum (dashed) bond dimensions for
    the dynamics shown in Fig.~\ref{fig:structured_dynamics}.}
  \label{fig:structured_bond}
\end{figure}
The dynamics of the same excitonic dimer system under this structured
vibrational bath is shown in Fig.~\ref{fig:structured_dynamics} using a time
step of $\Delta t = \SI{9.07}{\femto\second}$. A truncation cutoff of
$\chi=10^{-9}$ was used. For this simulation, we did not truncate memory. We
note that there are significant differences between
Fig.~\ref{fig:structured_dynamics} and Fig.~\ref{fig:exciton}. It has been shown
that increasing the bath reorganization energy, $\lambda$, or the characteristic
cutoff frequency, $\gamma$, leads to a suppression of the electronic
oscillations in this system~\cite{Ishizaki2009}. The first interesting feature
is that despite having a large reorganization energy of $\lambda =
  \SI{217.66}{\per\centi\meter}$ and comprising almost solely of high frequency
modes, the dynamics of the system coupled to the discrete bath demonstrate
clearer oscillations than those shown when coupled to the Drude bath of less
than half the reorganization energy ($\lambda = \SI{100}{\per\centi\meter}$).
Additionally, the first peak is at $\sim\SI{175}{\femto\second}$ for the
discrete bath, whereas it happens at $\sim\SI{125}{\femto\second}$ for the
$\lambda = \SI{100}{\per\centi\meter}$ Drude bath. Second, there are uneven and
irregular properties to the dynamics after $\sim\SI{300}{\femto\second}$.
Similar, yet more prominent, irregular features have already been
reported~\cite{Bose2020,Kundu2020a}. These features are due to the absence of a
continuous manifold of vibrational states guaranteed by a model spectral
density. Since, for simplicity, we are dealing with two isolated
bacteriochlorophyll molecules, the vibrational modes are discrete. This is, of
course, not the whole story. The presence of a protein scaffolding would lead to
a bath common to all the monomers with many ro-translational modes. In
principle, we should be able to include such common baths with AP-TNPI as well.
In Fig.~\ref{fig:structured_bond}, we plot the average bond dimension as a
function of time. Even though these are full memory calculations, the cost
clearly does not scale exponentially. It is important to note that the exact
characteristics of the time-evolution of the bond-dimension, even within memory,
is extremely parameter-dependent.

\subsection{Multilevel systems}
For our first multilevel system example, we extend the exciton model to a system
larger than a dimer. The Fenna-Matthew-Olson (FMO) complex is one of the most
widely studied light harvesting pigment-protein complexes.  A lot of
theoretical~\cite{olbrichTimeDependentAtomisticView2010,olbrichTheorySimulationEnvironmental2011,rengerNormalModeAnalysis2012,adolphsHowProteinsTrigger2006}
and experimental~\cite{Ratsep2011} effort has been expended in order to
accurately determine the site energies, electronic couplings and vibrational
spectral density of the complex. Additionally, various theoretical methods have
been used to study the correspond exciton transfer
dynamics.~\cite{nalbachRoleDiscreteMolecular2012,taoSemiclassicalDescriptionElectronic2010,ishizakiTheoreticalExaminationQuantum2009,mulvihillSimulatingEnergyTransfer2021}.
Here, as a simple demonstration, we use the well-known FMO model characterized
by the 7-site system Hamiltonian~\cite{adolphsHowProteinsTrigger2006}. This
model was studied using HEOM
by~\citet{ishizakiTheoreticalExaminationQuantum2009} using site-local Drude
baths describing the local
vibrations.\cite{ishizakiTheoreticalExaminationQuantum2009}

The Hamiltonian describing the bath and its interaction with the system is given
by Eq.~\ref{eq:exiton_Hb}. It has the same structure as that of the excitonic
dimer, except with seven sites, corresponding to the unique states in the single
excitation subspace. As with the other multi-bath models, all baths are taken to
be described by the same spectral density. A site-local Drude spectral density
is used with $\gamma=\SI{106.18}{\per\centi\meter}$ and
$\lambda=\SI{35}{\per\centi\meter}$.\cite{ishizakiTheoreticalExaminationQuantum2009}

\begin{figure}[h]
  \subfloat[Dynamics of the exciton subsequent to excitation of the first monomer.]{\includegraphics{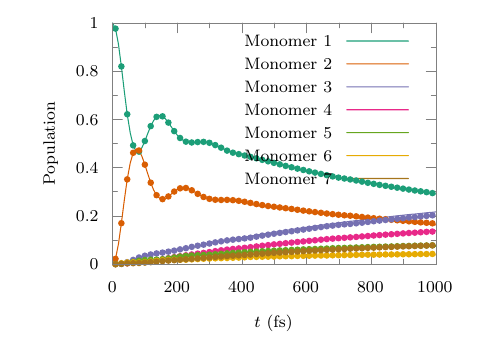}}
  
  \subfloat[Dynamics of the exciton subsequent to excitation of the sixth
    monomer.]{\includegraphics{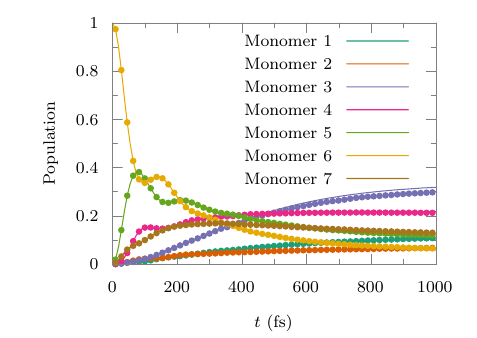}}
  
  \caption{Dynamics of 7-site FMO model. Markers: MTNPI results ($L=37$). Lines:
    AP-TNPI results ($L=27$).}\label{fig:fmo}
\end{figure}

\begin{figure}
  \includegraphics{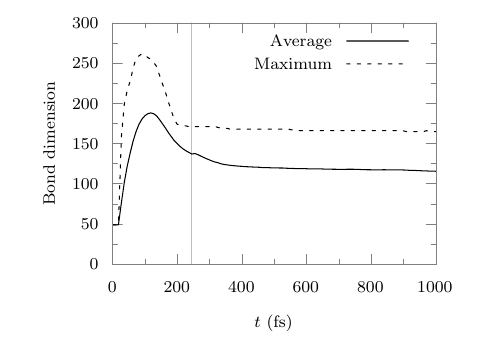}
  \caption{Growth of average (solid) and maximum (dashed) bond dimension with time for FMO. Gray vertical line indicates the memory span.}\label{fig:fmobond}
\end{figure}

In Fig.~\ref{fig:fmo}, we demonstrate the dynamics of the exciton, at
$T=\SI{300}{\kelvin}$, when the first and the sixth monomers are excited,
respectively, reproducing the results previously
observed~\cite{ishizakiTheoreticalExaminationQuantum2009}. The IF MPO is
obtained as a product of individual IF MPOs corresponding to the local baths of
each bacteriochlorophyll monomer as described in Appendix~\ref{app:if_local}.
The dynamics was converged with a timestep of $\Delta t =
  \SI{9.07}{\femto\second}$ and a memory length of $L=27$. The cutoff was
converged at $\chi = 10^{-8}$. For the MTNPI simulation, $L = 37$.  We find
quite good agreement between the MTNPI and full AP-TNPI results. In
Fig.~\ref{fig:fmobond},  we show the average and maximum bond dimension as a
function of time. It is interesting to note that in this case, the bond
dimensions starts decreasing even before the memory length is spanned. This is
in sharp contrast to most of the other cases, where the bond dimension increases
upto memory length and remains approximately constant thereafter. Between
Fig.~\ref{fig:fmobond} and Fig.~\ref{fig:structured_bond}, it is evident that it
is not simple to articulate a unifying rule to predict the evolution of bond
dimensions within or beyond memory. It needs to be investigated on a
case-by-case basis.

Finally, we explore the dynamics of a multilevel systems with a global solvent.
As an application, let us consider a molecular wire for charge transport. The
system is described by a tight-binding Hamiltonian with $N$ sites:
\begin{align}
  \hat{H}_{0} & = \sum_{1\le j\le D}\epsilon_{j}\dyad{\sigma_{j}} - \hbar V\sum_{1\le j<D}\left(\dyad{\sigma_{j}}{\sigma_{j+1}} + \dyad{\sigma_{j+1}}{\sigma_{j}}\right)\label{eq:mol_wire}
\end{align}
The sites are separated by unit distance. Therefore, the site locations are given
by $\hat{s}\ket{\sigma_{j}} = (j-1)\ket{\sigma_{j}}$ for all $1\le j\le D$. The process is
started with the $j=1$ site being populated. The bath is initially equilibrated
with this $j=1$ site. The presence of such a global bath prevents use of more
recent methods like modular path integrals~\cite{Makri2018a,Makri2018} that are
based on the separability of local environments. Following Lambert and
Makri~\cite{Lambert2012b}, we choose $V = 0.025$ and $\epsilon_{1} = 1$. For all other
sites, $1<j\le D$, $\epsilon_{j} = 0$. The bath is characterized by the Ohmic spectral
density defined in Eq.~\ref{eq:spectral} with $\omega_{c}=4$, $\xi=0.12$.

\begin{figure}
  \subfloat[Cold bath, $\beta=2$. Converged memory length $L=24$. MTNPI iteration
    with
    $L=320$.]{\includegraphics{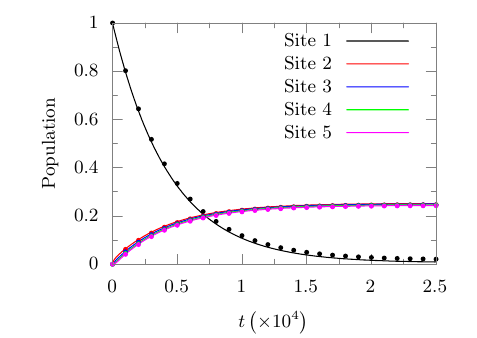}}
  
  \subfloat[Hot bath, $\beta=0.01$. Converged memory length $L=10$. MTNPI iteration
    with
    $L=30$.]{\includegraphics{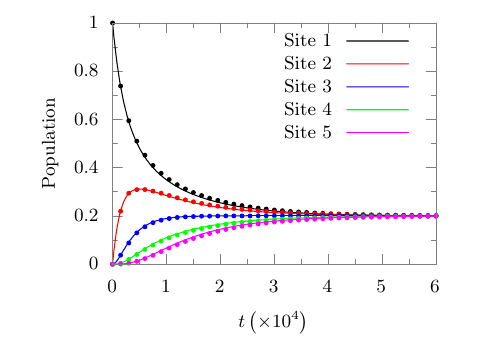}}
  \caption{Population dynamics of a molecular wire of chain of length $D=5$
    described by Eq.~\ref{eq:mol_wire}. Lines represent the MTNPI iteration
    scheme. Points represent the rigorously correct
    algorithm.}\label{fig:mol_wire_5}
\end{figure}
In Fig.~\ref{fig:mol_wire_5}, we show the dynamics of the population for a
molecular wire of length, $D=5$ at two different inverse temperatures, $\beta =
  2$ (Fig.~\ref{fig:mol_wire_5}~(a)) and $\beta=0.01$
(Fig.~\ref{fig:mol_wire_5}~(b)). A cutoff of $\chi=10^{-7}$ was used for the low
temperature calculation, and for the high temperature simulation $\chi =
  10^{-12}$ was used. For both cases, a time-step of $\Delta t = 0.2$ was used.
Because of the structure of this problem, even with such a small cutoff, the
maximum bond dimension for both the temperatures was stable at 25, which is also
the size of the forward-backward Hilbert space. The average bond dimension
decreases to 13.96 for the low temperature case and $15.18$ for the high
temperature case. The bond dimensions in this problem are significantly smaller
than those encountered in some of the two-level system problems, as well as the
FMO example. There are two factors responsible for this. First, the global
solvent causes a significant decrease in the probability amplitude for transfer
between sites as the distance between them increases. For example, a direct
transfer between the first site and the fifth site is extremely improbable
because of the high reorganization energies involved. The TNPI filtration scheme
seems to take advantage of this and can drastically reduce the bond dimensions
of the PA MPSs. Second, because of the definition of the system states,
$\hat{s}\ket{\sigma_{j}} = (j-1)\ket{\sigma_{j}}$, and the coupling to the bath,
there are only 9 unique values of $\Delta s$ as compared to the 25 possible
values of the forward-backward state. This leads to a greatly compressed
representation of the analytic IF MPO.

The higher temperature bath tends to lead to a sequential transfer of the excess
charge, whereas the colder bath leads to an almost simultaneous transfer of
population onto all the acceptor sites. The MTNPI scheme is quite accurate: we
get the basic features of the dynamics for the low temperature case, and
quantitative agreement with the correct dynamics for the high temperature
simulation. Though it requires much higher memory lengths to converge, the net
time required for the full simulation is orders of magnitude less than the
canonical algorithm for iteration. This is because, in the MTNPI scheme, the
full complexity of the longer memory only affects the within-memory portion of
the simulation. Thereafter, MTNPI abandons the PA tensor and just performs
tensor contractions over indices of dimension $D^2$, as shown in
Eqs.~\ref{eq:MTNPIa} and~\ref{eq:MTNPIb}. In contrast, for the canonical and
rigorously correct scheme of iteration, the computational complexity remains
approximately constant once in the iterative regime of the dynamics. This
becomes a significant bottleneck when the iterative regime consists of hundreds
of thousands of time steps.

The effective asymmetry in the low temperature case possibly leads the spurious
MTNPI memory to increase very rapidly. For the high temperature simulation, the
asymmetry is not important on the scale of the thermal energy, and consequently,
the spurious MTNPI memory is not as long compared to the converged memory of the
full AP-TNPI calculation. Because MTNPI propagates the reduced density matrix of
the system from time $t$ to $t+L\Delta t$ with an effective augmented
propagator, it assumes that the bath and the system are uncorrelated at time
$t$. We hypothesize that this implicit assumption performs poorly when the
asymmetry of the system is important. This would be consistent with what was
observed in QCPI~\cite{Walters2016}. Though this limitation can probably be
dealt with in a computationally efficient manner by augmenting MTNPI with the
ability to update the augmented propagator to be used in a dynamically
consistent manner, it is not important for the purposes of this example.

\section{Conclusion}\label{sec:conclusion}
System-solvent decomposition is a ubiquitous approach to quantum dynamics in
condensed phase where the quantum effects are primarily limited to a
low-dimensional subspace. We have presented a tensor network approach to
simulating such systems using the quasi-adiabatic propagators and Feynman-Vernon
influence functionals. This brings a novel perspective and decomposition to a
well-established problem.

Tensor networks utilize certain symmetries in the tensors to obtain highly
efficient ``factorized'' representations. The observation that the entanglement
between time-points cannot span temporal lengths greater than that of the
non-Markovian memory associated with the bath response function suggests
extremely compact matrix-product state representation of the augmented
propagator. Because the influence functional can be written as a product of
commuting terms, many representations are possible. Here, we have derived an
analytic form for the IF MPO that automatically provides a highly efficient and
compressed representation without having to resort to numerical optimization
techniques. There are well-established techniques for SVD-based and variational
optimizations of MPSs that lead to further automatic compression of the PA MPS.
This enables us to simulate problems with long non-Markovian memory lengths and
with large system sizes, without sacrificing the rigorous nature that
characterizes QuAPI-based methods. The structure of AP-TNPI suggests an
extremely simple and powerful approximate scheme for computing the reduced
density matrix beyond the memory length. This MTNPI method reduces the
computational complexity of each step to the cost of a single tensor contraction
involving indices with dimension $D^{2}$. Despite introducing additional memory,
using it is especially useful when the total time of propagation far exceeds the
memory time. Because AP-TNPI already allows us to access long memory lengths, it
is not very difficult to span the increased MTNPI memory as well.

After testing the new AP-TNPI framework against benchmark spin-boson
simulations, we applied it to study two realistic applications. First, we used a
combination of direct dynamics and rate theory to understand the physics of
charge transfer in ferrocene-ferrocenium solvated in hexane. These calculations
were performed on extremely modest computers and the time-scales that we
obtained were consistent with those found in previous QCPI simulations. As a
second example of a two-level system, we used AP-TNPI to simulate the exciton
dynamics of a chlorophyll dimer with both a Drude bath and a bath of discrete
vibrations. We reproduced results obtained by \citet{Ishizaki2009}. The bath of
discrete vibrations leaves some unusual features on the dynamics as compared to
the standard model spectral densities. Apart from two-level systems, we also
applied AP-TNPI to simulate the dynamics of an exciton in the 7-site model of
FMO coupled to local vibrational dissipative environments and a charge transfer
in a molecular wire coupled to global solvents. In the molecular wire examples,
the computational efficiency of MTNPI could truly be leveraged because the
memory length required to be spanned was much shorter than the total time of
propagation. Though quite accurate otherwise, it seems to perform comparatively
less efficiently when the asymmetry in the system is important. We have
postulated a possible solution, that we would explore in the future to make
general applications with MTNPI even more accurate. Both the many-site examples
show very interesting performance characteristics that have been discussed.
Further research is required to properly characterize and understand these
differences.

We have demonstrated the wide-ranging applicability of AP-TNPI to a variety of
problems using only modest computational resources. From our explorations, it
seems that AP-TNPI performs uniformly better than traditional QuAPI. We have
also demonstrated the applicability of this method to cases that are extremely
challenging even for BSPI. The very recently developed SMatPI and AP-TNPI are
substantially different methods and may have complementary benefits. A more
thorough investigation would be required to comment on the comparative merits of
the two. Future applications would focus on simulation of spectra in various
systems such as photosynthetic complexes and an analysis of the effect of a
structured spectral density of the bath on these spectra. While the current
examples of the excitonic dimer incorporate a local bath, it is possible to
incorporate effects of a global ro-translational solvent. This, along with
atomistic simulations with classical trajectories, would allow us to evaluate
the accuracy of linear response in modeling the protein backbones.

Since AP-TNPI is still a path integral technique, many developments in path
integrals involving improved reference propagators~\cite{Banerjee2013a} and
harmonic back reaction~\cite{Wang2019} to cheaply include anharmonic effects via
classical trajectories can be transparently implemented in AP-TNPI. It is
well-known that the use of classical trajectory-based reference propagators also
lead to an increase of the converged time-steps and decrease the effective
memory length~\cite{Banerjee2013a,Walters2016}. Additionally, a different and
extremely promising avenue of exploration relates to application of AP-TNPI to
extended systems. Simulations of energy and charge transfer along long molecular
chains and aggregates are of particular interest. We have already demonstrated
applications of the AP-TNPI framework introduced here to small multi-state
systems interacting with a global solvent. However, for longer, extended systems
with short-range interaction between sites, the structure of AP-TNPI can also be
augmented to decompose the system as a further optimization.

Future development would follow two broad avenues: incorporation of anharmonic
and atomistic environments using classical trajectories, and development of
spatial decomposition using tensor networks to efficiently work with extended
systems containing relatively short-ranged interactions. We believe that AP-TNPI
is a potentially viable framework for developing methods to study the dynamics
of several important condensed phase problems.

\section*{Acknowledgments}
We thank Nancy Makri for introducing us to the world of quantum dynamics in
general and real time path integrals in particular. A.\,B. acknowledges the
support of the Computational Chemical Center: Chemistry in Solution and at
Interfaces funded by the US Department of Energy under Award No. DE-SC0019394.
P.\,W. acknowledges the Miller Institute for Basic Research in Science for
funding.

\appendix
\section{Influence Functional for Localized Baths}\label{app:if_local}
In Sec.~\ref{sec:method}, we discussed the method primarily for a harmonic bath
bilinearly coupled to the ``position'' operator, $\hat{s}$, of the system.
However, the influence functionals can be used in a straightforward manner with
multiple baths that couple to different system operators. Here we summarize the
basic equations~\cite{Strumpfer2009,Nalbach2010} and changes to the algorithm
required to make AP-TNPI work with the more general bath.

Suppose after QuAPI splitting, the Hamiltonian for independent baths acting
through separate system operators is given as
\begin{align}
  \hat{H}_{b} & = \sum_{j,b} \frac{p_{jb}^{2}}{2m_{jb}} + \frac{1}{2}m_{jb}\omega_{jb}^{2}\left(x_{jb} - \frac{c_{jb}\hat{V}_{j}}{m_{jb}\omega_{jb}^{2}}\right)^{2}.
\end{align}
Here for each of the system operators indexed by $j$, there is an independent
bath with modes indexed by $b$. The frequency and coupling of the
$b$\textsuperscript{th} mode of the $j$\textsuperscript{th} bath is
$\omega_{jb}$ and $c_{jb}$ respectively. The $j$\textsuperscript{th} bath
interacts with the system through $\hat{V}_{j}$. For path integral calculations,
we want to work in a representation of the system where $\hat{V}_{j}$ is
diagonal for all $j$.

The independent baths are specified by the spectral densities
\begin{align}
  J_{j}(\omega) & = \frac{\pi}{2}\sum_{b}\frac{c_{jb}^{2}}{m_{jb}\omega_{jb}}\delta\left(\omega - \omega_{jb}\right).
\end{align}
which lead to separate $\eta$-coefficients for each of the baths. The
$\eta$-coefficient for the $j$\textsuperscript{th} bath is denoted by $\eta^{j}$.
These independent, local baths feature most prominently in descriptions of the
local vibrations of monomers in exciton transfer complexes and chromophores.

The only change to the dynamics happens through a change in the influence
functional, which becomes a product of all the influence functionals
corresponding to the different baths as discussed in Ref.~\cite{Nalbach2010}.
More specifically, Eq.~\ref{eq:influence_functional} would now be replaced by
\begin{align}
  F\left[\left\{s_{n}^{\pm}\right\}\right] & = \prod_{j}\exp\left(-\frac{1}{\hbar}\sum_{0\le k\le N}\Delta V_{jk}\right.\nonumber                                                      \\
                                           & \left.\sum_{0\le k'\le k}\left(\Re\left(\eta^{j}_{kk'}\right)\Delta V_{jk'} + 2i\Im\left(\eta^{j}_{kk'}\right)\bar{V}_{jk'}\right)\right)
\end{align}
where $\Delta V_{jk} = V_{j}\left(s_{k}^{+}\right) -
  V_{j}\left(s_{k}^{-}\right)$ and $\bar{V}_{jk} =
  \frac{V_{j}\left(s_{k}^{+}\right) + V_{j}\left(s_{k}^{-}\right)}{2}$. Here, we
have summarized Eq.~(A5) in the appendix of Ref.~\cite{Strumpfer2009}, or
equivalently, Eq.~(7) of Ref.~\cite{Nalbach2010} using the average and
difference coordinates and highlighting the product nature of the total
influence functional. Note that the main notational difference from Eq.~(7) of
Ref.~\cite{Nalbach2010} is that we use the $\hat{V}$ operator to encode the
system-solvent coupling for the multibath case instead of simply $\hat{s}$. In
fact, \citet{Nalbach2010} go further and discuss the influence functionals for
correlated baths on different sites, which have not been used here.

Therefore, instead of having one IF MPO, now there would be as many IF MPOs as
there are independent baths. The total IF MPO is defined to be the product of
all the individual ones. It is possible to compress these IF MPOs analytically
into one single compressed MPO, thereby reducing numerical error, and
computational complexity. However, for simplicity this is form is not described
here.

\bibliography{paper}
% \bibliography{/home/amartya/Documents/MendeleyDesktopLibrary/library}
\end{document}